\documentclass[pdftex,twocolumn,epjc3]{svjour3}          

\RequirePackage[T1]{fontenc}

\smartqed  

\RequirePackage{graphicx}
\RequirePackage{mathptmx}      
\RequirePackage{flushend}
\RequirePackage[numbers,sort&compress]{natbib}
\RequirePackage[colorlinks,citecolor=blue,urlcolor=blue,linkcolor=blue]{hyperref}

\usepackage{calc}
\usepackage{hyperref}
\usepackage{subfig}
\usepackage{float}
\usepackage{amsmath}
\usepackage{amsmath,bm}
\usepackage{amssymb}
\usepackage{latexsym}
\usepackage{color}
\usepackage{mciteplus}
\usepackage{fancyhdr}
\usepackage{booktabs}
\usepackage{colortbl}

\definecolor{Black}{gray}{0}
\definecolor{Gray}{gray}{0.85}
\definecolor{LightGray}{gray}{0.93}
\definecolor{LightGreen}{rgb}{0.88, 1, 0.88}
\definecolor{LightGreen}{rgb}{0.7, 1, .7}
\definecolor{LightCyan}{rgb}{0.88,1,1}
\definecolor{LightRed}{rgb}{1, 0.25, 0.25}
\definecolor{LightRed}{rgb}{1, 0.6, 0.65}
\definecolor{LightRed}{rgb}{1, 0.85, 0.85}
\definecolor{LightYellow}{rgb}{1, 1, 0.85}
\definecolor{LightYellow}{rgb}{1, 1, 0.85}
\definecolor{LightBlue}{rgb}{0.87, 0.94, 1}
\definecolor{White}{rgb}{1,1,1}
\definecolor{LightBlue}{rgb}{0.8, 1, 1}
\definecolor{white}{gray}{1}

\journalname{Eur. Phys. J. C}

\newcommand{\newc}{\newcommand*}

\long\def\begincomment#1\endcomment{%
        \begingroup\sf\baselineskip12pt#1\endgroup}

\newc{\etal}{\textrm{et al.}} 
\newc{\eg}{\textrm{e.g.}} 
\newc{\ie}{\textrm{i.e.}}
\newc{\etc}{\textrm{etc}}
\newc\vs{\textrm{vs.}}
\newc{\cl}{\rm {C.L.}}

\newc{\ev}{\ensuremath{\,\mathrm{eV}}}
\newc{\kev}{\ensuremath{\,\mathrm{keV}}}
\newc{\mev}{\ensuremath{\,\mathrm{MeV}}}
\newc{\gev}{\ensuremath{\,\mathrm{GeV}}}
\newc{\tev}{\ensuremath{\,\mathrm{TeV}}}
\newc{\MeV}{\mev} 
\newc{\TeV}{\tev}
\newc{\invpb}{\ensuremath{/\text{pb}}}
\newc{\invfb}{\ensuremath{\,\textrm{fb}^{-1}}}
\newc\nb{\ensuremath{\,\mathrm{nb}}} \newc\pb{\ensuremath{\,\mathrm{pb}}} \newc\fb{\ensuremath{\,\mathrm{fb}}}
\newc\pc{\ensuremath{\,\mathrm{pc}}}
\newc\kpc{\ensuremath{\,\mathrm{kpc}}}
\newc\mpc{\ensuremath{\,\mathrm{Mpc}}}
\newc\ps{\ensuremath{\,\mathrm{ps}}} 
\newc\cmeter{\ensuremath{\,\mathrm{cm}}} 
\newc\meter{\ensuremath{\,\mathrm{m}}} 
\newc\kmeter{\ensuremath{\,\mathrm{km}}}
\newc\second{\ensuremath{\,\mathrm{s}}}
\newc\msecond{\ensuremath{\,\mathrm{ms}}}
\newc\nsecond{\ensuremath{\,\mathrm{ns}}}
\newc\psecond{\ensuremath{\,\mathrm{ps}}}

\newc{\chisqmin}{\ensuremath{\chi^2_{\mathrm{min}}}}
\newc{\Delchisq}{\ensuremath{\Delta\chi^2}}
\newc{\chisq}{\ensuremath{\chi^2}}
\newc{\like}{\ensuremath{\mathcal{L}}}

\newc\lsim{\ensuremath{\mathrel{\rlap{\lower4pt\hbox{\hskip1pt$\sim$}}\raise1pt\hbox{$<$}}}}
\newc\gsim{\ensuremath{\mathrel{\rlap{\lower4pt\hbox{\hskip1pt$\sim$}}\raise1pt\hbox{$>$}}}}
\newc{\VEV}[1]{\ensuremath{\langle #1 \rangle}}
\newc{\dl}{\ensuremath{\stackrel{\leftarrow}{D}}}
\newc{\dr}{\ensuremath{\stackrel{\rightarrow}{D}}}

\newc{\bcenter}{\begin{center}}    \newc{\ecenter}{\end{center}}
\newc{\bfl}{\begin{flushleft}}    \newc{\efl}{\end{flushleft}}
\newc{\bfr}{\begin{flushright}}    \newc{\efr}{\end{flushright}}

\newc{\bi}{\begin{itemize}}
\newc{\ei}{\end{itemize}}
\newc{\bed}{\begin{description}}
\newc{\eed}{\end{description}}
\newc{\ben}{\begin{enumerate}}
\newc{\een}{\end{enumerate}}

\newc{\be}{\begin{equation}}
\newc{\ee}{\end{equation}}
\newc{\bea}{\begin{eqnarray}}
\newc{\eea}{\end{eqnarray}}
\newc{\bfle}{\begin{flalign}}
\newc{\efle}{\end{flalign}}
\newc{\ra}{\rightarrow}

\newc{\alphas}{\ensuremath{\alpha_s}}
\newc{\alphatwo}{\ensuremath{\alpha_2}}
\newc{\alphaone}{\ensuremath{\alpha_1}}
\newc{\alphai}[1]{\ensuremath{\alpha_{#1}}}
\newc{\alphaem}{\ensuremath{\alpha_{\mathrm{em}}}}
\newc{\alphaeff}{\ensuremath{\alpha_{\mathrm{eff}}}}
\newc{\sineff}{\ensuremath{\sin \theta_{\mathrm{eff}}}}
\newc{\sinsqeff}{\ensuremath{\sin^2 \theta_{\mathrm{eff}}}}
\newc{\dalphahad}{\ensuremath{\Delta \alpha_{\mathrm{had}}}}
\newc{\yt}{\ensuremath{h_t}} \newc{\yb}{\ensuremath{h_b}} \newc{\ytau}{\ensuremath{h_{\tau}}}
\newc\mz{\ensuremath{M_Z}} 
\newc\mw{\ensuremath{m_W}}
\newc\mZ{\mz}        \newc\mW{\mw}
\newc\mhsm{\ensuremath{ m_{H_{\mathrm{SM}}}}}
\newc{\mtop}{\ensuremath{ m_t}}               \newc{\mtpole}{\ensuremath{ M_t}}
\newc{\mbottom}{\ensuremath{ m_b}} 
\newc{\mtau}{\ensuremath{ m_{\tau}}}
\newc{\mt}{\mtpole}
\newc{\mb}{\mbottom} 
\newc{\rtwogg}{\ensuremath{R_{h_2}(\gamma\gamma)}}
\newc{\rtwozz}{\ensuremath{R_{h_2}(ZZ)}}
\newc{\ronegg}{\ensuremath{R_{h_1}(\gamma\gamma)}}
\newc{\ronezz}{\ensuremath{R_{h_1}(ZZ)}}
\newc{\rsiggg}{\ensuremath{R_{h_\textrm{sig}}(\gamma\gamma)}}
\newc{\rsigzz}{\ensuremath{R_{h_\textrm{sig}}(ZZ)}}
\newc{\llbar}{\ensuremath{\ell\bar{\ell}}}
\newc{\tauptaum}{\ensuremath{ \tau^+\tau^-}}
\newc{\qqbar}{\ensuremath{ q\bar{q}}} \newc{\ppbar}{\ensuremath{ p\bar{p}}}
\newc{\bbbar}{\ensuremath{ b\bar{b}}} \newc{\ttbar}{\ensuremath{ t\bar{t}}}
\newc{\ffbar}{\ensuremath{ f\bar{f}}} \newc{\tautaubar}{\ensuremath{ \tau\bar{\tau}}}

\newc{\mchi}{\ensuremath{m_\neutone}}
\newc{\squark}{\ensuremath{\tilde{q}}}
\newc{\slepton}{\ensuremath{\tilde{l}}}
\newc{\gluino}{\ensuremath{\tilde{g}}} 
\newc{\mgluino}{\ensuremath{{m_{\gluino}}}}
\newc{\wino}{\ensuremath{\tilde{W}}} 
\newc{\mwino}{\ensuremath{{m_{\wino}}}}
\newc{\tone}{\ensuremath{{\tilde{t}_1}}}
\newc{\bone}{\ensuremath{{\tilde{b}_1}}}
\newc{\Hone}{\ensuremath{{\tilde{H}_{1}}}}
\newc{\Htwo}{\ensuremath{{\tilde{H}_{2}}}}
\newc{\Hhtwo}{\ensuremath{{H_{2}}}}
\newc{\qli}{\ensuremath{{\tilde{Q}_{i}}}}
\newc{\uri}{\ensuremath{{\tilde{u}_{i}}}}
\newc{\dri}{\ensuremath{{\tilde{d}_{i}}}}
\newc{\lli}{\ensuremath{{\tilde{L}_{i}}}}
\newc{\eri}{\ensuremath{{\tilde{e}_{i}}}}

\newc{\sthw}{\ensuremath{ \sin\theta_W}}              \newc{\cthw}{\ensuremath{\cos\theta_W}}
\newc{\tanthw}{\ensuremath{ \tan\theta_W}}              \newc{\cotthw}{\ensuremath{\cot\theta_W}}
\newc{\ssqthw}{\ensuremath{\sin^2 \theta_W}}
\newc{\msbar}{\ensuremath{\overline{MS}}} \newc{\drbar}{\ensuremath{\overline{DR}}}
\newc{\mtmtsmmsbar}{\ensuremath{ m_t(m_t)^{\msbar}_{{\mathrm{SM}}}}}
\newc{\mtmtsmdrbar}{\ensuremath{ m_t(m_t)^{\drbar}_{{\mathrm{SM}}}}}
\newc{\mtmtmssmdrbar}{\ensuremath{ m_t(m_t)^{\drbar}_{{\mathrm{SUSY}}}}}
\newc{\mbmbmsbar}{\ensuremath{ m_b(m_b)^{\msbar} }}
\newc{\mbmbsmmsbar}{\ensuremath{ m_b(m_b)^{\msbar}_{{\mathrm{SM}}}}}
\newc{\mbmzsmmsbar}{\ensuremath{ m_b(\mz)^{\msbar}_{{\mathrm{SM}}}}}
\newc{\mbmzsmdrbar}{\ensuremath{ m_b(\mz)^{\drbar}_{{\mathrm{SM}}}}}
\newc{\mbmzmssmdrbar}{\ensuremath{ m_b(\mz)^{\drbar}_{{\mathrm{SUSY}}}}}
\newc{\mtaumzsmmsbar}{\ensuremath{ m_{\tau}(\mz)^{\msbar}_{{\mathrm{SM}}}}}
\newc{\mtaumzsmdrbar}{\ensuremath{ m_{\tau}(\mz)^{\drbar}_{{\mathrm{SM}}}}}
\newc{\mtaumzmssmdrbar}{\ensuremath{ m_{\tau}(\mz)^{\drbar}_{{\mathrm{SUSY}}}}}
\newc{\alphasmzms}{\ensuremath{\alpha_s(M_Z)^{\overline{MS}}}}
\newc{\alphaimzms}[1]{\ensuremath{\alpha_{#1}(M_Z)^{\overline{MS}}}}
\newc{\alphaemmz}{\ensuremath{\alpha_{\mathrm{em}}(M_Z)^{\overline{MS}}}}

\newc{\mzero}{\ensuremath{{m_0}}}
\newc{\mhalf}{\ensuremath{ m_{1/2}}}
\newc{\tanb}{\ensuremath{\tan\beta}}
\newc{\azero}{\ensuremath{ A_0}}
\newc{\signmu}{\ensuremath{\rm{sgn}\,\mu}}
\newc{\atau}{\ensuremath{{A_{\tau}}}}
\newc{\mueff}{\ensuremath{\mu_{\rm{eff}}}}
\newc{\lam}{\ensuremath{{\lambda}}}
\newc{\kap}{\ensuremath{{\kappa}}}
\newc{\alam}{\ensuremath{{A_{\lambda}}}}
\newc{\akap}{\ensuremath{{A_{\kappa}}}}
\newc{\hs}{\ensuremath{ H_s}}      
\newc{\mhs}{\ensuremath{ m_{H_s}}} 
\newc{\mgut}{\ensuremath{ M_{\rm GUT}}}
\newc{\gut}{\ensuremath{{\rm GUT}}}
\newc{\mplanck}{\ensuremath{ M_{\rm P}}}      \newc{\mpl}{\ensuremath{ M_{\rm Pl}}}
\newc{\msusy}{\ensuremath{ M_{\rm SUSY}}}      \newc{\ms}{\ensuremath{ M_{\rm S}}}
 \newc{\hu}{\ensuremath{ H_u}}       \newc{\hd}{\ensuremath{ H_d}}
 \newc{\mhu}{\ensuremath{ m_{H_u}}}       \newc{\mhd}{\ensuremath{ m_{H_d}}}
 \newc{\mhuew}{\ensuremath{ m^{\ast}_{H_u}}}       \newc{\mhdew}{\ensuremath{ m^{\ast}_{H_d}}}
 \newc{\mhuewsq}{\ensuremath{ m^{\ast\, 2}_{H_u}}}       \newc{\mhdewsq}{\ensuremath{ m^{\ast\, 2}_{H_d}}}
 \newc{\mhl}{\ensuremath{m_\hl}} 
 \newc{\mhone}{\ensuremath{m_{h_1}}} 
 \newc{\mhtwo}{\ensuremath{m_{h_2}}} 
 \newc{\mhi}{\ensuremath{m_{\tilde{h}}}} 
 \newc{\mul}{\ensuremath{m_{\tilde{u}_L}}} 
 \newc{\mbone}{\ensuremath{m_{\tilde{b}_1}}}  
 \newc{\mtone}{\ensuremath{m_{\tilde{t}_1}}} 
 \newc{\ma}{\ensuremath{m_A}} 
 \newc{\mH}{\ensuremath{m_H}} 
 \newc{\maone}{\ensuremath{m_{a_1}}} 
 \newc{\matwo}{\ensuremath{m_{a_2}}}
 \newc{\hone}{\ensuremath{h_1}}
 \newc{\htwo}{\ensuremath{h_2}}
 \newc{\aone}{\ensuremath{a_1}}
 \newc{\atwo}{\ensuremath{a_2}}
 \newc{\mqthree}{\ensuremath{m_{\tilde{Q}_3}^2}}
 \newc{\muthree}{\ensuremath{m_{\tilde{u}_3}^2}}
 \newc{\mqli}{\ensuremath{m_{\tilde{Q}_{i}}}}
 \newc{\muri}{\ensuremath{m_{\tilde{u}_{i}}}}
 \newc{\mdri}{\ensuremath{m_{\tilde{d}_{i}}}}
 \newc{\mlli}{\ensuremath{m_{\tilde{L}_{i}}}}
 \newc{\meri}{\ensuremath{m_{\tilde{e}_{i}}}}
 \newc{\ts}{\ensuremath{T_{SUSY}}}

\newc{\sigsip}{\ensuremath{\sigma^{\rm SI}_{p}}}	\newc{\sigsin}{\ensuremath{\sigma^{\rm SI}_{n}}}
\newc{\sigsdp}{\ensuremath{\sigma^{\rm SD}_{p}}}	\newc{\sigsdn}{\ensuremath{\sigma^{\rm SD}_{n}}}
\newc{\sigsi}{\ensuremath{\sigma^{\rm SI}}}	\newc{\sigsd}{\ensuremath{\sigma^{\rm SD}}}
\newc{\abund}{\ensuremath{ \Omega h^2}}
\newc{\omegadm}{\ensuremath{ \Omega_{{\rm DM}}}}     \newc{\abunddm}{\ensuremath{ \Omega_{{\rm DM}} h^2}} 
\newc{\omegam}{\ensuremath{ \Omega_{{\rm m}}}}       \newc{\abundm}{\ensuremath{ \Omega_{{\rm m}} h^2}}
\newc{\omegab}{\ensuremath{ \Omega_{{\rm b}}}}	\newc{\abundb}{\ensuremath{ \Omega_{{\rm b}} h^2}}
\newc{\omegatot}{\ensuremath{ \Omega_{{\rm TOT}}}}
\newc{\omegacdm}{\ensuremath{ \Omega_{{\rm CDM}}}}   \newc{\abundcdm}{\ensuremath{ \Omega_{{\rm CDM}} h^2}}
\newc{\omegalambda}{\ensuremath{ \Omega_{\Lambda}}} \newc{\abundlambda}{\ensuremath{ \Omega_{\Lambda} h^2}}
\newc{\omegarad}{\ensuremath{ \Omega_{{\rm rad}}}}  \newc{\abundrad}{\ensuremath{ \Omega_{{\rm rad}} h^2}}
\newc{\rhocrit}{\ensuremath{ \rho_{\rm crit}}}
\newc{\rhochi}{\ensuremath{ \rho_{\chi}}}
\newc{\abunchi}{\ensuremath{\Omega_\chi h^2}}
\newc{\abundlsp}{\ensuremath{\Omega_{\rm LSP}h^2}}

\newc{\amu}{\ensuremath{ a_{\mu}}}        \newc{\amususy}{\ensuremath{ a_{\mu}^{\mathrm{SUSY}}}}
\newc{\amuexpt}{\ensuremath{ a_{\mu}^{\mathrm{expt}}}}        \newc{\amusm}{\ensuremath{ a_{\mu}^{\mathrm{SM}}}}
\newc\deltaamu{\ensuremath{\Delta a_{\mu}}} \newc{\deltaamususy}{\ensuremath{\delta a_{\mu}^{\mathrm{SUSY}}}}
\newc\gmtwo{\ensuremath{ (g-2)_{\mu}}} 
\newc{\deltagmtwomususy}{\ensuremath{\delta\left(g-2\right)_{\mu}^{\mathrm{SUSY}}}}
\newc{\deltagmtwomu}{\ensuremath{\delta\left(g-2\right)_{\mu}}}
\newc\BR{\ensuremath{\rm BR}}
\newc\bsgamma{\ensuremath{ b\rightarrow s \gamma }}
\newc\bxsgamma{\ensuremath{\overline{B}\rightarrow X_{s}\gamma}}
\newc\brbsgamma{\ensuremath{\BR\left(\bsgamma\right)}}
\newc\brbxsgamma{\ensuremath{\BR\left(\bxsgamma\right)}}
\newc\bsmumu{\ensuremath{B_s\to\mu^+\mu^-}}
\newc\brbsmumu{\ensuremath{\BR\left(B_s\to\mu^+\mu^-\right)}}
\newc\bdmmumu{\ensuremath{\overline{B}_d\to\mu^+\mu^-}}
\newc\bbbarmix{\ensuremath{\overline{B}_s\mbox{-}B_s}}      
\newc\delmbs{\ensuremath{\Delta M_{B_s}}}
\newc{\butaunu}{\ensuremath{B_u \rightarrow \tau \nu}}
\newc{\brbutaunu}{\ensuremath{\BR\left(B_u \rightarrow \tau \nu\right)}}

\newcommand*{\reftable}[1]{Table~\ref{#1}}

\newcommand*{\reffig}[1]{Fig.~\ref{#1}}
 
        \newcommand*{\refeq}[1]{Eq.~(\ref{#1})}
     \newcommand*{\refsec}[1]{Sec.~\ref{#1}}

\newcommand*{\neutone}{\ensuremath{\chi^0_1}}

\newcommand*{\flavio}{{\tt flavio}}

\let\oldcite\cite
\renewcommand*{\cite}{~\oldcite}

\newcommand*{\hl}{\ensuremath{h}}

\begin{document}

\title{Implications for New Physics in $b\to s \mu\mu$ transitions after recent measurements by Belle and LHCb}


\author{Kamila Kowalska,\thanksref{e1}
        Dinesh Kumar\and \thanksref{e2}
        Enrico Maria Sessolo\thanksref{e3} 
}

\thankstext{e1}{e-mail: kamila.kowalska@ncbj.gov.pl}
\thankstext{e2}{e-mail: dinesh.kumar@ncbj.gov.pl}
\thankstext{e3}{e-mail: enrico.sessolo@ncbj.gov.pl}

\institute{National Centre for Nuclear Research, Pasteura 7, 02-093 Warsaw, Poland}

\date{Received: date / Accepted: date}

\maketitle


\begin{abstract}
We present a Bayesian analysis of the implications for new physics
in semileptonic $b\to s$ transitions after including new measurements of $R_K$ at LHCb and new determinations of 
$R_{K^*}$ and $R_{K^{*+}}$ at Belle.
We perform global fits with 1, 2, 4, and 8 input Wilson coefficients, plus one CKM nuisance parameter to take into account 
uncertainties that are not factorizable. We infer the 68\% and 95.4\% credibility regions of the marginalized posterior probability density for all scenarios
and perform comparisons of models in pairs by calculating the Bayes factor 
given a common data set.
We then proceed to analyzing a few well-known BSM models that can provide a high energy framework for the EFT analysis.
These include the exchange of a heavy $Z^{'}$ boson in models with heavy vector-like fermions and a scalar field,
and a model with scalar leptoquarks. We provide predictions for the BSM couplings and expected mass values.
\end{abstract}

\section{Introduction\label{sec:intro}}
The LHCb Collaboration has recently presented a new measurement of the observable $R_K$, the ratio of the branching fraction of $B$-meson decay into a kaon and muons, over the decay to a kaon and 
electrons, from the combined analyses of the Run~1 and partially of Run~2 data set\cite{Aaij:2019wad}, which reads 
\begin{equation}
R_K = 0.846^{+0.060+0.016}_{-0.054-0.014}\,.
\end{equation}
This new measurement of $R_K$ is compatible with the Standard Model (SM) prediction at $2.5\,\sigma$ significance. At the same time, the Belle Collaboration has presented new results for the observable $R_{K^*}$ in $B^0$-meson decays, as well as the first measurement of its counterpart  $R_{K^{*+}}$ in $B^+$ decays\cite{Abdesselam:2019wac}. 
These new results, listed in \reftable{tab:belle_result}, are consistent with the SM at $1\,\sigma$, mainly due to large experimental uncertainties.

\begin{table}[b]
\centering
\begin{tabular}{|c|c|c|}
\hline 
$q^2$ in GeV$^2$ & $R_{K^*} $ &$R_{K^{*+}} $ \\ 
\hline
$[0.045,1.1]$ & $0.46^{+0.55}_{-0.27} \pm 0.07$ & $0.62^{+0.60}_{-0.36} \pm 0.10$\\
\hline 
$[1.1,6.0]$ & $1.06^{+0.63}_{-0.38} \pm 0.13$ & $0.72^{+0.99}_{-0.44} \pm 0.18$\\
\hline
$[0.045,1.1]$ & $1.12^{+0.61}_{-0.36} \pm 0.10$ & $1.40^{+1.99}_{-0.68} \pm 0.11$\\
\hline
\end{tabular}
\caption{$R_{K^*}$ and $R_{K^{*+}}$ results from Belle\cite{Abdesselam:2019wac}.} 
\label{tab:belle_result} 
\end{table}

The rare decays of $B$ mesons are known to provide fertile testing ground for physics beyond the Standard Model (BSM), 
as in the SM they are highly suppressed by the smallness of the Cabibbo-Kobayashi-Maskawa (CKM) matrix elements and/or by helicity. 
While for many observables an anomalous determination does not necessarily imply the presence of New Physics (NP), as 
the QCD uncertainties can be sizable, ratios like 
$R_K$ and $R_{K^*}$ provide fairly clean probes, with parametric uncertainties that cancel out to high precision. 
Additionally, a deviation from the SM in these observables would imply a violation of lepton-flavor universality (LFUV), a purely BSM phenomenon.

Over the last few years the rare decays of $B$-mesons involving $b \to s ll$ interactions have attracted a lot of attention for the search of BSM physics. The update of the 
LHCb results had been long awaited, as Run~1 determinations of $R_K$ and $R_{K^*}$\cite{Aaij:2014ora,Aaij:2017vbb} both featured a $2$--$3\,\sigma$ deficit with respect to the SM. They were
also part of a broader set of anomalous measurements in rare semileptonic $B$ decays obtained at the LHC and Belle\cite{Aaij:2014pli,Aaij:2015oid,Aaij:2015esa,Wehle:2016yoi,Aaboud:2018krd,Aaij:2018gwm}, 
which involved $b\to s$ transitions and muons in the final state, and  
which in global statistical analyses\cite{Altmannshofer:2014rta,Altmannshofer:2017fio,Capdevila:2017bsm,Altmannshofer:2017yso,
DAmico:2017mtc,Alok:2017sui,Ciuchini:2017mik,Hurth:2014vma,Hurth:2016fbr,Chobanova:2017ghn,Hurth:2017hxg,Arbey:2018ics} had been shown to favor strongly 
the presence of a few NP operators over the SM.
Remarkably, the post-Run~1 global fits presented in Refs.\cite{Altmannshofer:2014rta,Altmannshofer:2017fio,Capdevila:2017bsm,Altmannshofer:2017yso,
DAmico:2017mtc,Alok:2017sui,Ciuchini:2017mik,Hurth:2014vma,Hurth:2016fbr,Chobanova:2017ghn,Hurth:2017hxg,Arbey:2018ics} 
often differed from one another in the choice of the full experimental data set and input parameters, and in the treatment of the parametric theoretical uncertainties, but they overall 
reached the same conclusions as to the high statistical significance of the NP effects, which, according to some studies, touched the $\sim 6\,\sigma$ level. 

References\cite{Altmannshofer:2014rta,Altmannshofer:2017fio,Capdevila:2017bsm,Altmannshofer:2017yso,
DAmico:2017mtc,Alok:2017sui,Ciuchini:2017mik,Hurth:2014vma,Hurth:2016fbr,Chobanova:2017ghn,Hurth:2017hxg,Arbey:2018ics}, however, did not differ from one another in 
their choice of adopted statistical framework, as all performed a frequentist, chi-squared or profile-likelihood based analysis, providing confidence intervals  
and the pull of the best-fit points from the SM. An alternative and complementary measure of the goodness of fit, 
best indicated for the comparison of competing models that can equally well explain the data, is furnished by Bayesian statistics. Specifically, one could compute the 
Bayesian evidence to quantify how well one given model agrees with the data, and the Bayes factor
to estimate which of the competing models is more likely to be the real one. With respect to the profile-likelihood technique to derive confidence intervals, 
Bayesian inference of the posterior probability density function (pdf) has the advantage of being less subject to the risk of under-coverage and, through the procedure of marginalization, 
incorporates in the computation of the credible regions effects that depend globally on the full parameter space of the model. The Bayesian approach has also the  
advantage of providing a well-defined prescription for the treatment of nuisance parameters, which contribute to the theoretical uncertainty of the fit.  
  
Thus, in light of the very recent measurements of $R_K$ and $R_{K^*}$ by LHCb and Belle, we think this is the time to perform a global Bayesian analysis of the BSM effects 
appearing in the combination of the new data with older results from the LHC and Belle. On the one hand the paper provides an update to the Run~1 global fits mentioned
above, and at the same time it follows the spirit of earlier Bayesian analyses of radiative $B$-mesons decays\cite{Beaujean:2013soa,Ghosh:2014awa}. 

We first consider model-independent fits to the global set of flavor observables, within the framework of the electroweak effective field theory (EFT), 
assuming as input the Wilson coefficients of 
four-fermion vector operators that were shown to be able to accommodate the observed data in Run~1 better than the SM.
We perform scans with 1, 2, 4, and 8 independent input parameters, plus one nuisance parameter, the $V_{cb}$ element of the CKM matrix, which takes into account 
uncertainties that are not factorizable. For each scan we infer the 68\% and 95.4\% credibility regions of the marginalized posterior 
pdf. We then compare models in pairs and provide the Bayes factor for a given set of data.
Additionally, we make contact with the frequentist approach by providing for each scan the best-fit point, its pull from the SM, and an estimate of the goodness of fit. 

In concomitance with our work, Refs.\cite{Alguero:2019ptt,Alok:2019ufo,Ciuchini:2019usw,Datta:2019zca,Aebischer:2019mlg} have recently appeared, 
most of which analyze the effect of incorporating the new LHCb and Belle data in a frequentist context. The part of our analysis based on the chi-squared distribution 
agrees with those studies, but we repeat that we focus in this article on 
the computation of the multi-dimensional Bayesian posterior pdf, and on the use of Bayes factors to discriminate among models.
  
In the second part of this paper we proceed to analyzing a few well known models that can provide a high energy framework for the EFT analysis.
Depending on the case and on which set of observables is included in the global fit, BSM interpretations at the tree level 
have involved the exchange of a heavy $Z^{'}$ boson 
(see Refs.\cite{Buras:2012jb,Gauld:2013qja,Altmannshofer:2014cfa,Buras:2013dea,Crivellin:2015mga,Chiang:2016qov,Bonilla:2017lsq,DiChiara:2017cjq} for early studies) or of a leptoquark (studies include Refs.\cite{Gripaios:2014tna,Becirevic:2015asa,Fajfer:2015ycq,Varzielas:2015iva,Hiller:2014yaa,Calibbi:2015kma,Bauer:2015knc,
Barbieri:2015yvd,Becirevic:2016yqi,Assad:2017iib,DiLuzio:2017vat,Becirevic:2017jtw,Becirevic:2018uab,Fornal:2018dqn,Blanke:2018sro}), 
both with non-universal and flavor-violating couplings to leptons and quarks. 
Moving one step further, such couplings can be generated assuming heavy vector-like (VL) fermions that can mix with the SM fermions\cite{Boucenna:2016wpr,Boucenna:2016qad}.
We perform a few global scans for some of the $Z'$ models with VL fermions, and for one leptoquark model consistent with the flavor anomalies.\footnote{A third possibility explored in the literature
is to generate the desired Wilson coefficients at the loop level with box diagrams, see e.g. Refs.\cite{Gripaios:2015gra,Arnan:2016cpy,Crivellin:2018yvo,Marzo:2019ldg}. 
These explanations generally require 
large Yukawa couplings. We do not explore explicit examples 
here and we remand the reader to the original papers.} 

The paper is organized as follows. In \refsec{sec:method}, we describe the Bayesian methodology used in our analysis. In \refsec{sec:eft}, we perform a model-independent global fit to the full set of $b\to s$ observables. Predictions for several extensions of the SM that can accommodate the observed anomalies are presented in Sec.~\ref{sec:mod}. Finally, we summarize our findings in Sec.~\ref{sec:summary}.

\section{Fit methodology\label{sec:method}}

For each model described by a set of input parameters we map out the regions of the parameter space that are in best agreement with all relevant
experimental constraints. To this end we use Bayesian statistics, whose main features we briefly summarize here.

In the Bayesian approach, for a theory described by some parameters $m$, experimental observables
$\xi(m)$ can be compared with data $d$ and a pdf $p(m|d)$, of the model parameters $m$, can be calculated through Bayes' Theorem. This reads
\begin{equation}
p(m|d)=\frac{p(d|\xi(m))\pi(m)}{p(d)}\, ,
\label{Bayesth}
\end{equation}
where the likelihood $p(d|\xi(m))\equiv\mathcal{L}(m)$ gives the probability density for obtaining $d$ from a
measurement of $\xi$ given a specific value of $m$, and the prior $\pi(m)$ parametrizes
assumptions about the theory prior to performing the measurement. The
evidence, $p(d)\equiv\mathcal{Z}$, is a function of the data that depends globally on the model's parameter space.  
As long as one considers only one model the evidence is a normalization constant, but it serves as a comparative measure for different models or scenarios.

Bayes' theorem provides an efficient and 
natural procedure for drawing inferences on a subset of $r$ variables in the parameter space, $\psi_{i=1,..,r}\subset m$.
One just needs to marginalize, or integrate, the posterior pdf over the remaining
parameters,
\begin{equation}
p(\psi_{i=1,..,r}|d)=\int p(m|d)d^{n-r}m\,,
\label{marginalization}
\end{equation}
where $n$ denotes the dimension of the full parameter space. 
In this work we will be interested in drawing the 68\% ($1\,\sigma$) and 95.4\% ($2\,\sigma$) marginalized 1-dimensional and 2-dimensional credible regions of the 
posterior pdf for each model under consideration.
We will also compare in pairs different models fitting to the same data set, to determine which one is favored by the data distribution.
We do this by computing the Bayes factor, defined as the ratio of evidences for two arbitrary models $\mathcal{M}_1$ and $\mathcal{M}_2$, 
i.e., $p(d)_{\mathcal{M}_1}/p(d)_{\mathcal{M}_2}$. We estimate the significance of Bayes factors according to Jeffrey's scale\cite{Jeffery:1998,Kass}. 

The central object in our statistical analysis is the likelihood
function, constructed using the following prescription. 
Given the set $m$ of input parameters, which can be, depending on the case, Wilson coefficients, particle masses, coupling constants, or other,  
the likelihood function is 
\begin{multline}
\mathcal{L}(m)=\\
 \exp\left\{-\frac{1}{2}\left[\mathcal{O}_{\textrm{th}}(m) -\mathcal{O}_{\textrm{exp}}\right]^T
  \left(\mathcal{C}^{\textrm{exp}} +\mathcal{C}^{\textrm{th}}\right)^{-1}\right.\\ 
\left. \left[\mathcal{O}_{\textrm{th}}(m) -\mathcal{O}_{\textrm{exp}} \right]\right\},
\end{multline}
where $\mathcal{O}_{\rm th}$ gives a vector of theoretical predictions of the observables of interest and $\mathcal{O}_{\rm exp}$ is the vector of the experimental measurements of 
those observables. We have taken into account the available experimental correlation, which is encoded in the matrix $\mathcal{C}^{\textrm{exp}}$. The experimental correlation is available in angular 
observables for $B \to K^* \mu \mu$\cite{Aaij:2015oid} and $B_s \to \phi \mu\mu$\cite{Aaij:2015esa}. 

The theoretical correlation is given by the matrix $\mathcal{C}^{\textrm{th}}$, which is computed using {\tt flavio}\cite{Straub:2018kue}, 
in which hadronic form factors from lattice QCD are implemented\cite{Horgan:2013hoa,Horgan:2015vla,Detmold:2016pkz,Meinel:2016grj,Bailey:2015dka,Bazavov:2016nty}. 
The theoretical uncertainties, including possible correlations, are estimated as the standard deviation of the values of the observables, calculated by 
taking $N$ random choices of all input parameters (form factors, bag parameters, decay constants, masses of the particles)  
according to their probability distribution\cite{Straub:2018kue}. In this procedure, the precision with which the standard deviation is known 
increases with the number of random points. We take $N = 2000$ random points, which corresponds to a $\sim 2\%$ precision on the theoretical error estimate. 
The $V_{cb}$ element of the CKM matrix is treated as a real 
nuisance parameter. We scan it together with the models' input parameters, following a Gaussian distribution 
around its central Particle Data Group (PDG) value\cite{Tanabashi:2018oca}, and adopting PDG uncertainties. 

The statistical analysis performed in this study takes into account a large set of experimental measurements involving $b\to s$ transitions. 
The full list of all observables included in the likelihood function can be found in \ref{app:obs}. 
In the following we summarize them briefly:
\begin{itemize}
\item $R_K$ and $R_{K^*}$ (\reftable{tab:data_B0_exp0})
\item $B^0\to K^{*0}\mu^+\mu^-$: CP-averaged angular observables $S_{i=3,4,5,7,8,9}$\cite{Altmannshofer:2008dz}, 
fraction of longitudinal polarization of the $K^{*0}$ meson $F_L$, and forward-backward asymmetry of the dimuon system $A_{FB}$ (alternatively, CP-averaged optimized $P_{i=1,4,5}'$ observables can be used), binned differential branching ratio $d\textrm{BR}/dq^2$ (Tables~\ref{tab:data_B0_exp1}--\ref{tab:data_B0_exp4})
\item  $B^+\to K^+\mu^+\mu^-$, $B^+\to K^{*+}\mu^+\mu^-$, $B^0\to K^{0}\mu^+\mu^-$: binned differential branching ratios $d\textrm{BR}/dq^2$ (Tables~\ref{tab:data_B0_exp5}--\ref{tab:data_B0_exp7})
\item $B_s^0\to \phi\mu^+\mu^-$: time- and CP-averaged angular observables $S_{i=3,4,7}$, 
time-averaged fraction of longitudinal polarization $F_L$, and differential branching ratio $d\textrm{BR}/dq^2$ (Tables~\ref{tab:data_B0_exp8}--\ref{tab:data_B0_exp9})
\item $\Lambda_b^0\to \Lambda\mu^+\mu^-$: binned forward-backward asymmetries and binned differential branching ratios (Tables~\ref{tab:data_B0_exp10}--\ref{tab:data_B0_exp11})
\item $B^+\to K^+\mu^+\mu^-$: binned forward-backward asymmetry $A_{FB}$ (\reftable{tab:data_B0_exp12})
\item $B^0\to K^{*0}e^+ e^-$: CP-averaged angular observables $P'_{4,5}$, binned longitudinal polarization fraction $F_L$ and binned differential branching ratio $d\textrm{BR}/dq^2$ (Tables~\ref{tab:data_B0_exp14}--\ref{tab:data_B0_exp15})
\item $B^+\to K^+e^+ e^-$: binned differential branching ratio (\reftable{tab:data_B0_exp17})
\item binned branching ratios $\textrm{BR}(B_s^0\to X_s\mu^+\mu^-)$ and $\textrm{BR}(B_s^0\to X_se^+e^-)$ (\reftable{tab:data_B0_exp13})
\item time-integrated branching ratio $\overline{\textrm{BR}}(B_s^0 \to \mu^+\mu^-)$ (\reftable{tab:data_B0_exp18}). 
\end{itemize}

\section{Effective field theory analysis\label{sec:eft}}

In the model-independent approach we adopt the weak EFT framework. 
The effective Hamiltonian for the $b \to s ll$ transition can be written as:
\be \label{heff}
\mathcal{H}_{eff}=-\frac{4G_F}{\sqrt{2}}V_{tb}V_{ts}^* \sum_{i,l}(C_i^l O_i^l + C_i^{' l} O_i^{' l}) + \textrm{H.c.}\,,
\ee
where $G_F$ is the Fermi constant and $V_{tb}$, $V_{ts}$ are elements of the CKM matrix. 
In \refeq{heff} 
the short-distance physics is encoded in the Wilson coefficients $C_i^{(\prime)l}$ after integrating out the heavy degrees of freedom, 
whereas the long-distance physics is described by the four-fermion dimension-six interaction operators $O_i^{(\prime) l}$, invariant under the SU(3)$_{\textrm{c}}\times$U(1)$_{\textrm{em}}$ 
gauge group. In this study we will assume the presence of NP in the following semi-leptonic operators:
\bea
O_9^l&=&\frac{e^2}{16\pi^2}(\bar{s}_L\gamma^\mu b_L)(\bar{l}\gamma_\mu l),\nonumber\\ 
O_9^{' l}&=&\frac{e^2}{16\pi^2}(\bar{s}_R\gamma^\mu b_R)(\bar{l}\gamma_\mu l),\nonumber\\
O_{10}^l&=&\frac{e^2}{16\pi^2}(\bar{s}_L\gamma^\mu b_L)(\bar{l}\gamma_\mu\gamma_5 l),\nonumber\\
O_{10}^{' l}&=&\frac{e^2}{16\pi^2}(\bar{s}_R\gamma^\mu b_R)(\bar{l}\gamma_\mu\gamma_5 l),
\eea
where the lepton $l$ can be an electron or a muon. 
We restrict ourselves to the analysis of CP-conserving NP effects, 
so that the Wilson coefficient are assumed to be real.

We do not consider here NP in scalar and pseudoscalar operators, $O_S^{(\prime)}$ and $O_P^{(\prime)}$, as they are severely constrained by the $B_s \to \mu^+ \mu^-$ measurement\cite{Alonso:2014csa,Altmannshofer:2017wqy}. 
Similarly, the electromagnetic dipole operator $O_7^{(\prime)}$ is tightly constrained by radiative decays\cite{Paul:2016urs}. 
The remaining dimension-six operators, chromomagnetic dipole operators, and four-quark operators, at the leading order can only contribute to the semi-leptonic decays through the mixing 
into semi-leptonic operators. All other BSM contributions enter at a higher order, so that it is safe to consider them as negligible for the purposes of this analysis.

The Wilson coefficients defined in~\refeq{heff} contain both the SM and NP 
contributions, which can be written as, e.g., $C_i^l = C_i^{\textrm{SM}} + C_i^{l,\textrm{NP}}$, where, again,  $l=e,\mu$. 
In the SM, the Wilson coefficients at the scale $\mu=4.2\gev$ are lepton-flavor universal and read:
\be\label{sm_wilson}
C^{\textrm{SM}}_9=4.27,\qquad C^{\textrm{SM}}_{10}=-4.17,\qquad C^{\prime \textrm{SM}}_{9,10}=0.
\ee
NP contributions to the primed operators can in principle be significant and as such can be considered  a smoking gun for BSM phenomena.

\begin{table}[t]
\centering
\begin{tabular}{|c|c|c|}
\hline 
Parameter & Range & Prior \\ 
\hline
\hline 
$C_9^{\mu}$ & $(-3,3)$ & Flat \\ 
\hline 
$C_9^{\mu}=-C_{10}^{\mu}$ & $(-3,3)$ & Flat \\ 
\hline 
$C_9^{\mu}$, $C_{10}^{\mu}$ & $(-3,3)$ & Flat \\ 
\hline 
$C_9^{\mu}$, $C_9^{\prime\mu}$ &  $(-3,3)$ & Flat \\ 
\hline 
$C_9^{\mu}$, $C_{10}^{\mu}$, $C_9^{\prime\mu}$, $C_{10}^{\prime\mu}$ &  $(-3,3)$ & Flat \\ 
\hline 
$C_9^{\mu}$, $C_{10}^{\mu}$, $C_9^{e}$, $C_{10}^{e}$ &  $(-3,3)$ & Flat \\ 
\hline 
$C_9^{\mu}$, $C_{9}^{\prime\mu}$, $C_9^{e}$, $C_{9}^{\prime e}$ &  $(-3,3)$ & Flat \\ 
\hline 
$C_9^{\mu}$, $C_{9}^{\prime\mu}$, $C_{10}^{\mu}$, $C_{10}^{\prime \mu}$ &  $(-3,3)$ & Flat \\ 
$C_9^{e}$, $C_{9}^{\prime e}$, $C_{10}^{e}$, $C_{10}^{\prime e}$ &   & \\
\hline
$m_{Z'}/g_X$ & 500--5000\gev & Log \\
$M_Q/\lam_Q$, $M_D/\lam_D$ &  0.1--500\tev  & Log \\
\hline
$m_{Z'}/g_X$ & 500--5000\gev & Log \\
$M_Q/\lam_Q$, $M_E/\lam_{E,2}$ &  0.1--500\tev  & Log \\
\hline 
\hline
Nuisance parameter & Central value, error ($\times 10^{-2}$)&  \\
\hline
\hline
CKM matrix element $V_{cb}$ & (4.22, 0.08)\cite{Tanabashi:2018oca} & Gaussian \\
\hline
\end{tabular}
\caption{Input parameters, their ranges, and prior distributions for the 10 scans we run in this study.} 
\label{tab:priors} 
\end{table}

We perform in this work 8 separate EFT scan fits to the data, each with a different combination of Wilson coefficients as input parameters.
We summarize their input ranges and prior distributions in the first 8 lines of \reftable{tab:priors}  
(here and in what follows we drop the superscript ``NP'' from the Wilson coefficients' names, 
but we always take as parameters the New Physics contribution).
In each determination we also simultaneously scan over the CKM 
matrix element $V_{cb}$, which we treat as a nuisance parameter for the Bayesian analysis.
We have checked with several preliminary scans that the latter is the only CKM matrix element that can substantially interfere 
with NP effects due to its large uncertainty, as was pointed out, e.g., in Ref.\cite{Altmannshofer:2014rta}.

We also perform one benchmark scan for the SM, in which all NP Wilson coefficients are set to zero, so the only varying input parameter is the nuisance parameter $V_{cb}$. In the Bayesian framework, the SM scan provides us with the value of evidence that allows to compare the SM with the considered NP scenarios through the Bayes factor. In the frequentist approach, the $\chi^2$ value for the best-fit point obtained in this scan is used as a reference value to calculate the NP pull from the SM. 

To get a better grip of the physics responsible for the shape of the parameter space favored in all considered scenarios, we will make use of  approximate formulas for $R_K$ and $R_{K^*}$, in which only the dominant linear  BSM  contributions are taken into account. With real Wilson coefficents and the polarization fraction of the $K^*$ meson set at $p=0.86$, they read\cite{Hiller:2014ula}
\bea
R_K&\approx & 1+\frac{2}{|C_9^{\textrm{SM}}|^2+|C_{10}^{\textrm{SM}}|^2}\left[C_9^{\textrm{SM}}\left(C_9^{\mu}+C_9^{\prime\mu}\right)\right.\nonumber\\
&+& \left.C_{10}^{\textrm{SM}}\left(C_{10}^{\mu}+C_{10}^{\prime\mu}\right)-\left(\mu\to e\right)\right],\label{RK_full1}\\
R_{K^*}&\approx & 1+\frac{2}{|C_9^{\textrm{SM}}|^2+|C_{10}^{\textrm{SM}}|^2}\left[C_9^{\textrm{SM}}C_9^{\mu}+C_{10}^{\textrm{SM}}C_{10}^{\mu}\right.\nonumber\\
&-&\left. 0.72 \left(C_9^{\textrm{SM}}C_9^{\prime\mu}+C_{10}^{\textrm{SM}}C_{10}^{\prime\mu}\right)-\left(\mu\to e\right)\right]\label{RK_full2}.
\eea
Substituting the SM numerical values in Eqs.~(\ref{RK_full1})-(\ref{RK_full2}), we further approximate the expressions as
\bea
R_K&\approx & 1+0.24\left(C_9^{\mu}-C_{10}^{\mu}+C_9^{\prime\mu}-C_{10}^{\prime\mu}\right)-\left(\mu\to e\right),\label{RK_form}\\
R_{K^*}&\approx & 1+0.24\left(C_9^{\mu}-C_{10}^{\mu}\right)-0.17\left(C_9^{\prime\mu}-C_{10}^{\prime\mu}\right)\nonumber\\
 &-&\left(\mu\to e\right).\label{RK_form1}
\eea

This section is dedicated to the discussion of the EFT fits.  
In \refsec{sec:mod}, we will instead investigate the implications of the new data for a few 
popular BSM models, involving a new gauge boson $Z'$ or a leptoquark, which are able to provide the favored parameter space regions for the Wilson coefficients. 
The input parameters, ranges and priors of two of those models, which will be described in the next section, 
occupy lines 9 and 10 of \reftable{tab:priors}.
 
\subsection{Discussion of results}

All the observables are calculated with \texttt{flavio}\cite{Straub:2018kue}, 
according to the procedure outlined in \refsec{sec:method}. 
In order to efficiently scan the multidimensional parameter space we used
\texttt{MultiNest v2.7}\cite{Feroz:2008xx} and \texttt{pyMultiNest}\cite{pyMulti} for sampling the parameter space and calculating the evidence. 
The 68\% ($1\,\sigma$) and 95.4\% ($2\,\sigma$) credible regions of the marginalized posterior pdf 
are computed and plotted with the public tool \texttt{Superplot}\cite{Fowlie:2016hew}.

\begin{figure*}[t]
\centering
\subfloat[]{%
\includegraphics[width=0.35\textwidth]{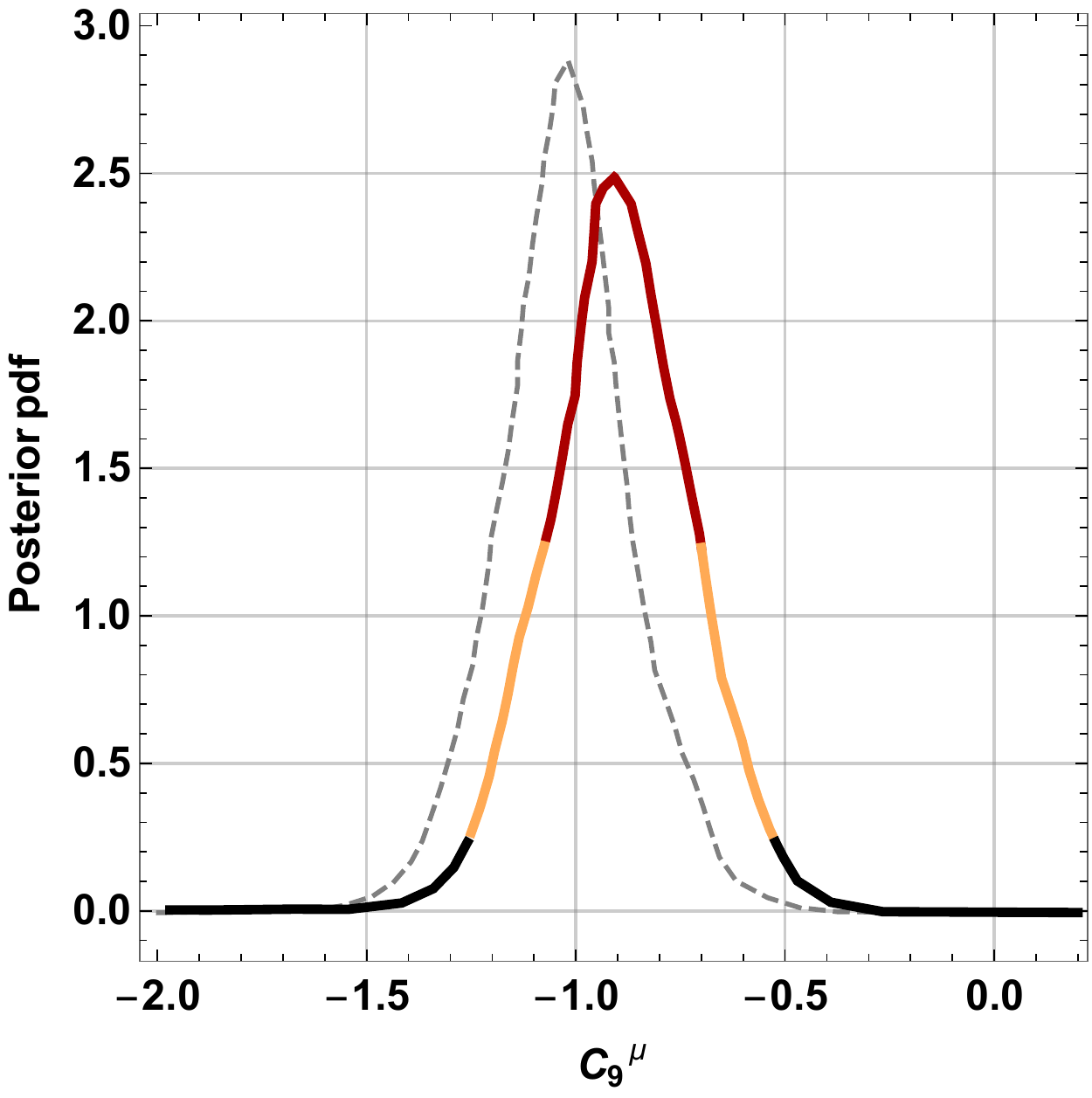}
}%
\hspace{0.02\textwidth}
\subfloat[]{%
\includegraphics[width=0.35\textwidth]{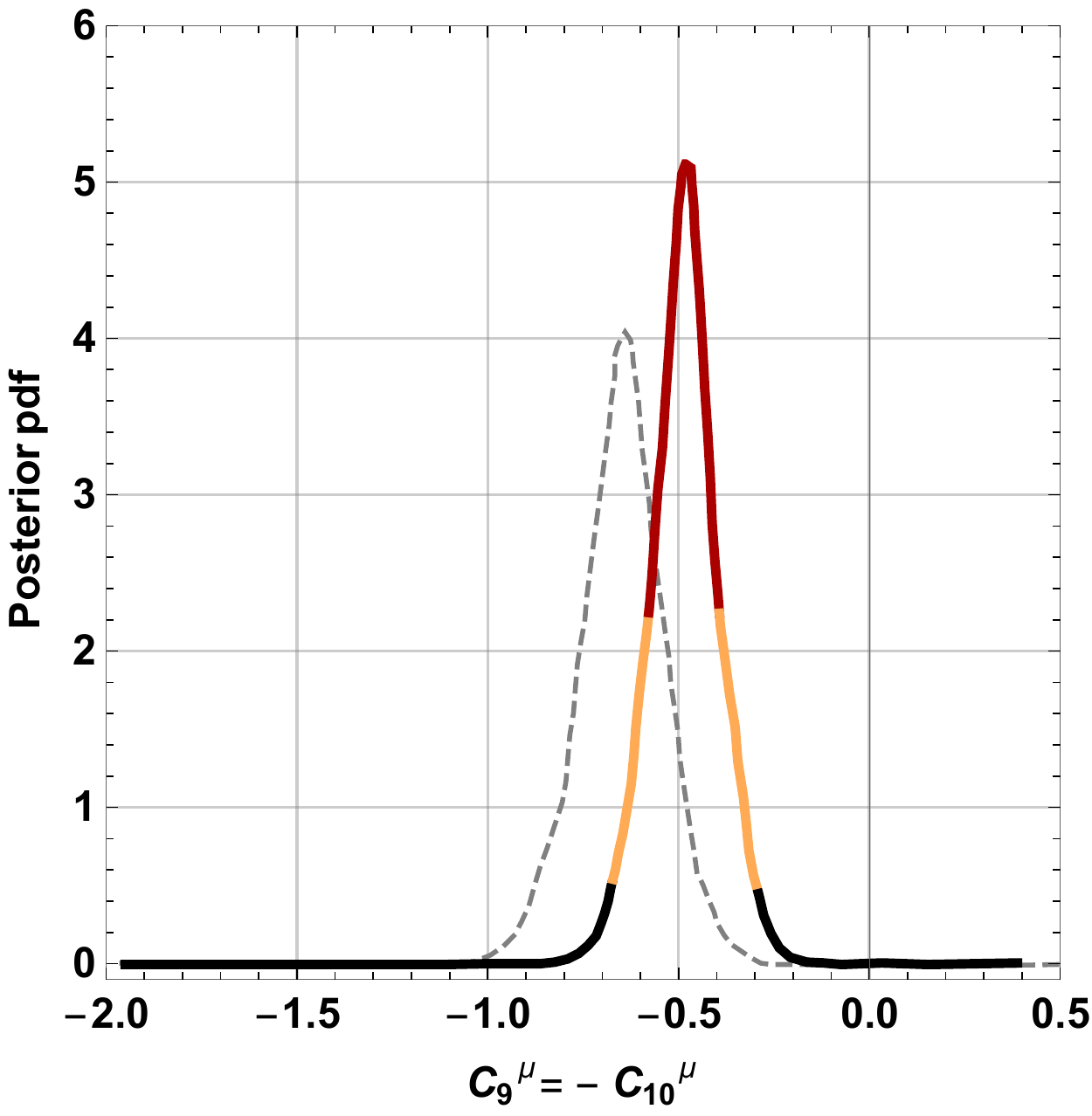}
}%
\caption{\footnotesize (a) 1-dimensional posterior pdf marginalized over the nuisance parameter for the scan in the input parameter $C_9^{\mu}$ (first row of \reftable{tab:priors}). Colors indicate the $1\,\sigma$ (red) and $2\,\sigma$ (orange) credible regions. 
The gray dashed line shows the posterior pdf corresponding to the data pre-LHCb Run~2. 
(b) Same as (a) for the scan parametrized by $C_9^{\mu}=-C_{10}^{\mu}$ (second row of \reftable{tab:priors}).
\label{fig:1par}}
\end{figure*}

In \reffig{fig:1par}(a) we show the posterior pdf for the model with a single nonzero Wilson coefficient $C_9^{\mu}$ (first row of \reftable{tab:priors}). In red and orange, $1\,\sigma$ and $2\,\sigma$ credible regions are indicated, respectively. For a comparison, in dashed gray we show the corresponding posterior pdf for the scan in which new Belle and LHCb results were not taken into account. The new data causes a shift of the favored $C_9^{\mu}$ towards lower values, the reason for which will be explained below. 
In \reffig{fig:1par}(b) we present the same quantities for the model with $C_9^{\mu}=-C_{10}^{\mu}$ (second row of \reftable{tab:priors}). 

\begin{figure*}[t]
\centering
\subfloat[]{%
\includegraphics[width=0.35\textwidth]{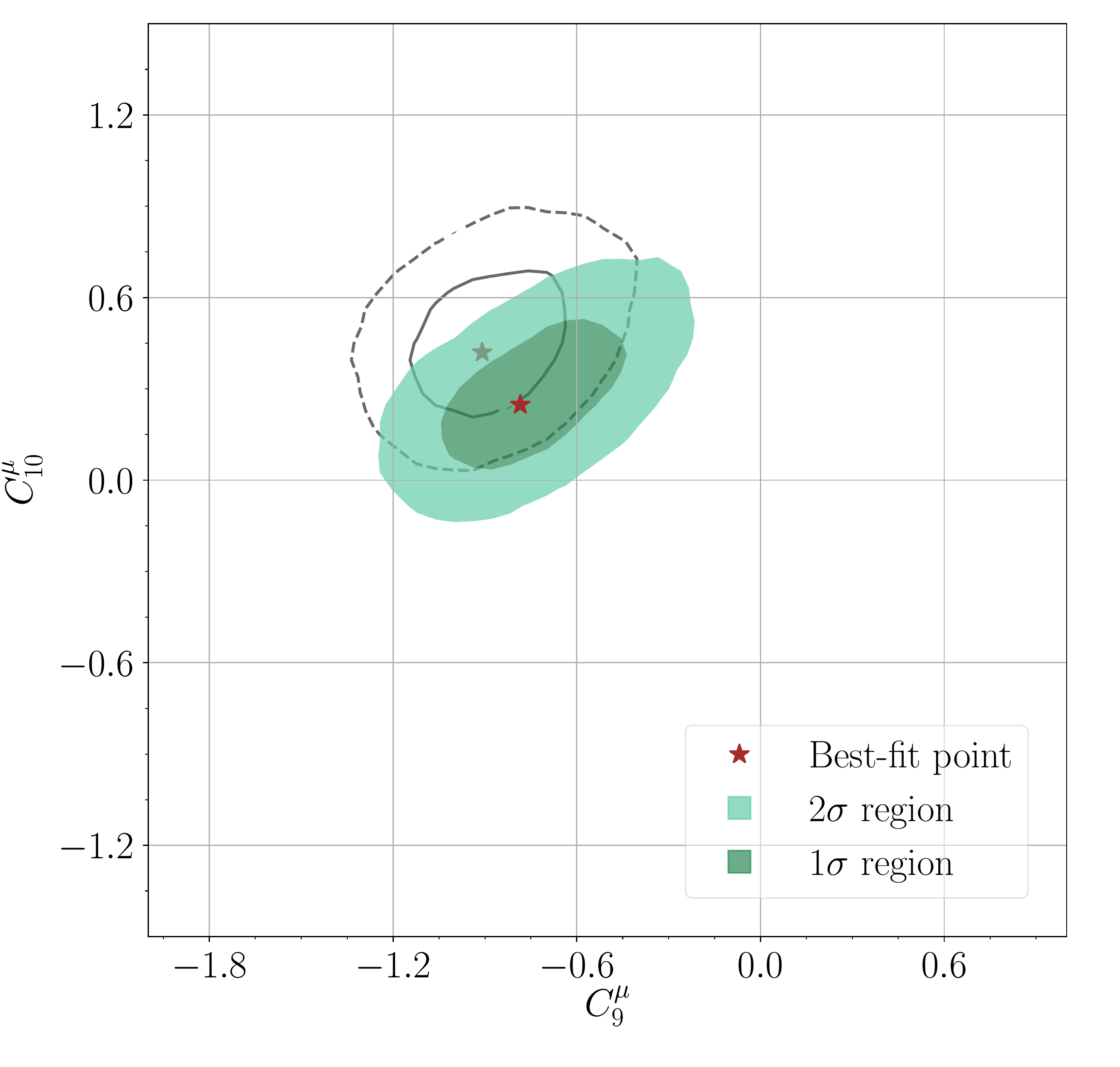}
}%
\hspace{0.02\textwidth}
\subfloat[]{%
\includegraphics[width=0.35\textwidth]{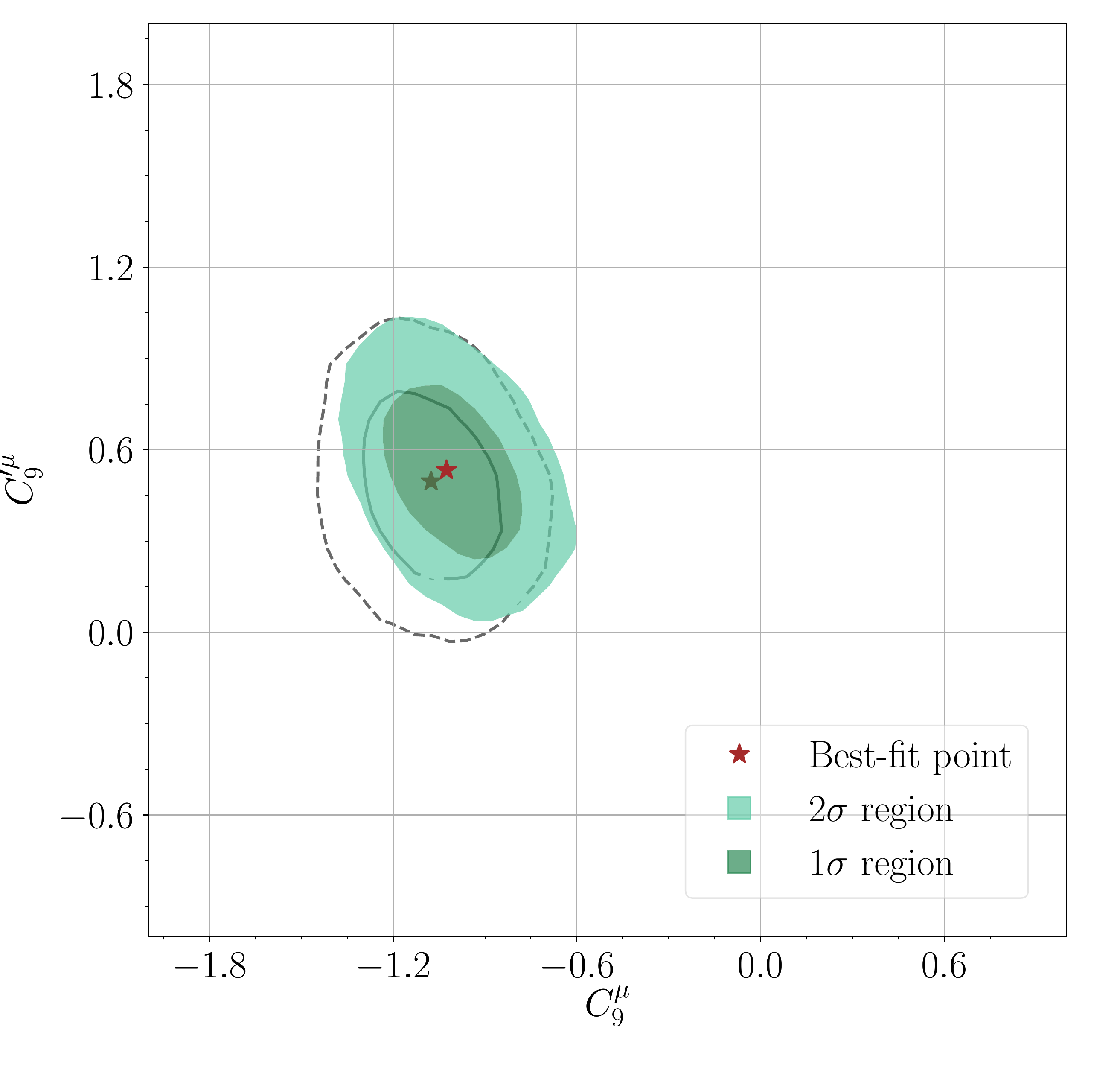}
}%
\caption{\footnotesize (a) In green, the $1\,\sigma$ (dark) and $2\,\sigma$ (light) credible regions of the posterior pdf for the scan in the input parameters $C_9^{\mu}$, $C_{10}^{\mu}$ 
(third row of \reftable{tab:priors}), marginalized over the nuisance parameter. The red star marks the position of the best-fit point. 
The gray solid (dashed) line shows the $1\,\sigma$ ($2\,\sigma$) credible region of the pdf corresponding to the data pre-LHCb Run~2. 
The associated best-fit point is also shown in gray. 
(b) Same as (a) for the scan parametrized by $C_9^{\mu}$, $C_9^{\prime\mu}$ (fourth row of \reftable{tab:priors}).
\label{fig:2pars}}
\end{figure*}

In \reffig{fig:2pars}(a) we show the $1\,\sigma$ and $2\,\sigma$ credible regions of the posterior pdf for the model in the third row of \reftable{tab:priors}, 
parametrized by $C_9^{\mu}$, $C_{10}^{\mu}$ and the nuisance parameter.
The posterior is compared to the one obtained with the previous data, pre LHCb Run~2. 
The overall value of $R_K$, higher than in the previous determination, has the effect of bringing the $2\,\sigma$ region closer to the axes origin.  
The modification of the posterior pdf is not large, but visible. In this case, in fact, one expects $R_K\approx R_{K^*}$, see \refeq{RK_form}, 
and a tension between the measurements of $R_K$ and $R_{K^*}$ arises. As a further consequence, the posterior pdf becomes narrower. 

One encounters a less substantial modification of the pdf in the case of the scan parametrized by $C_9^{\mu}$, $C_9^{\prime\mu}$ (fourth row of \reftable{tab:priors}), 
for which the credible regions are shown in \reffig{fig:2pars}(b). The overall effect appears to be a very slight detachment of the $2\,\sigma$ region 
from the $C_9^{\prime\mu}=0$ axis, which is once more confirmed by \refeq{RK_form}: in this case $R_K$ and $R_{K^*}$ can be fitted separately with a positive $C_9^{\prime\mu}$ so the new experimental measurements do not affect much the posterior pdf. 

A fit to the new data with 4 input NP parameters, 
$C_9^{\mu}$, $C_{10}^{\mu}$, $C_9^{\prime\mu}$, $C_{10}^{\prime\mu}$ (fifth row of \reftable{tab:priors})
shows that the introduction of $C_{10}^{\prime\mu}$ as a free parameter 
leads to an interesting interference with $C_9^{\prime\mu}$. The latter can be made thus comfortably consistent with zero, at the price of introducing 
a substantial negative value of $C_{10}^{\prime\mu}$, see \refeq{RK_form}. 

\begin{figure*}[t]
\centering
\subfloat[]{%
\includegraphics[width=0.35\textwidth]{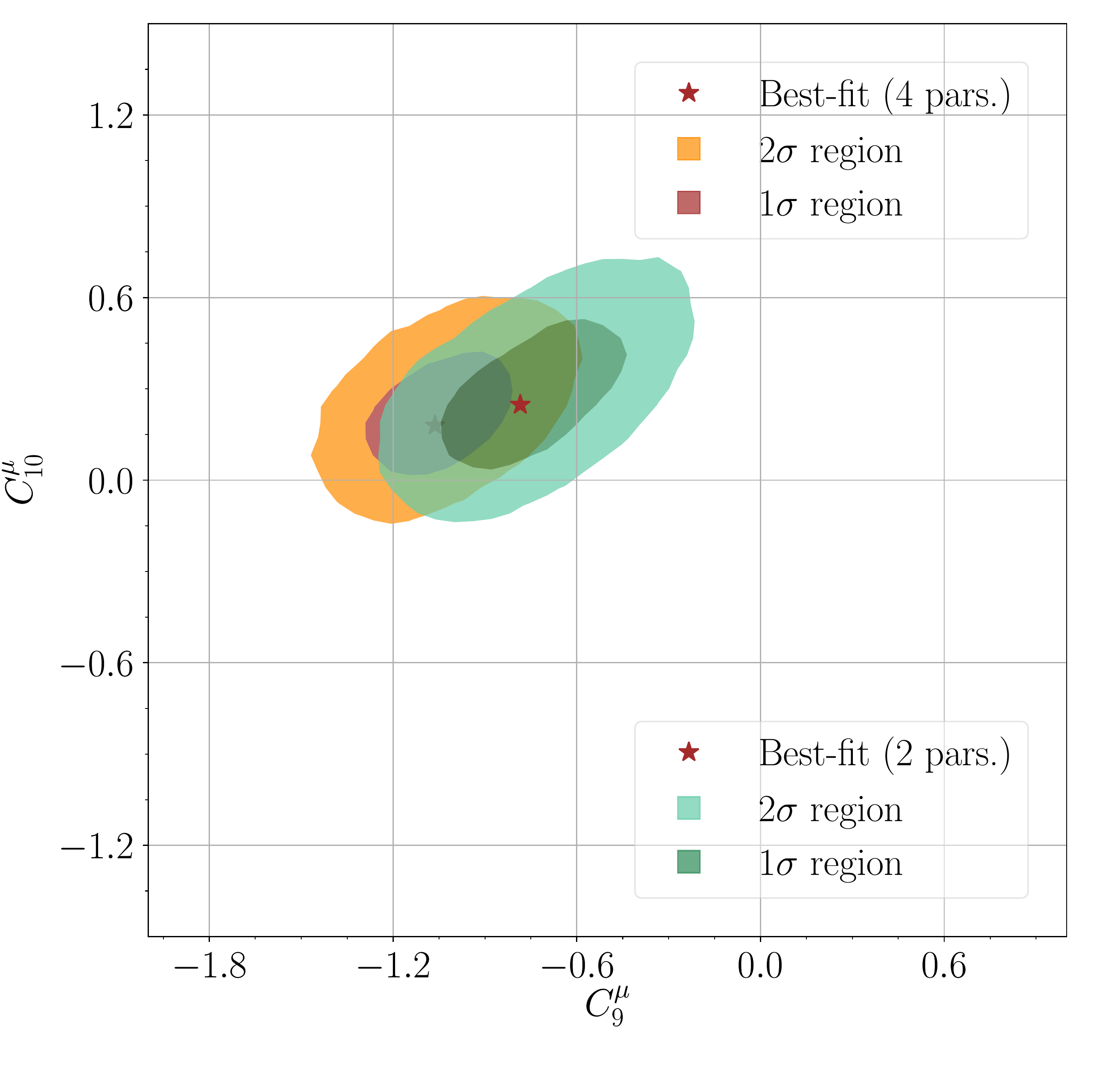}
}%
\hspace{0.02\textwidth}
\subfloat[]{%
\includegraphics[width=0.35\textwidth]{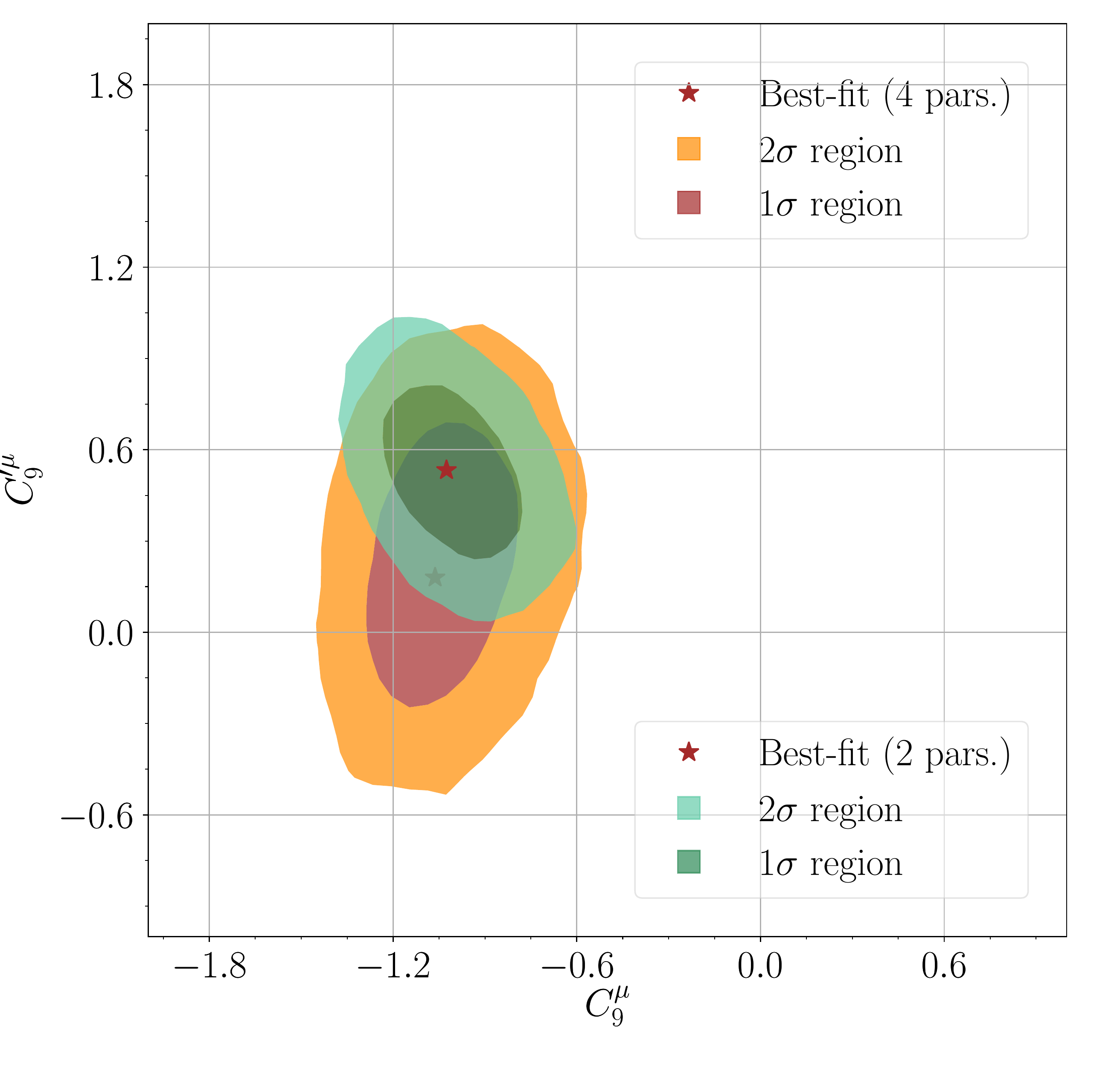}
}%
\\
\subfloat[]{%
\includegraphics[width=0.35\textwidth]{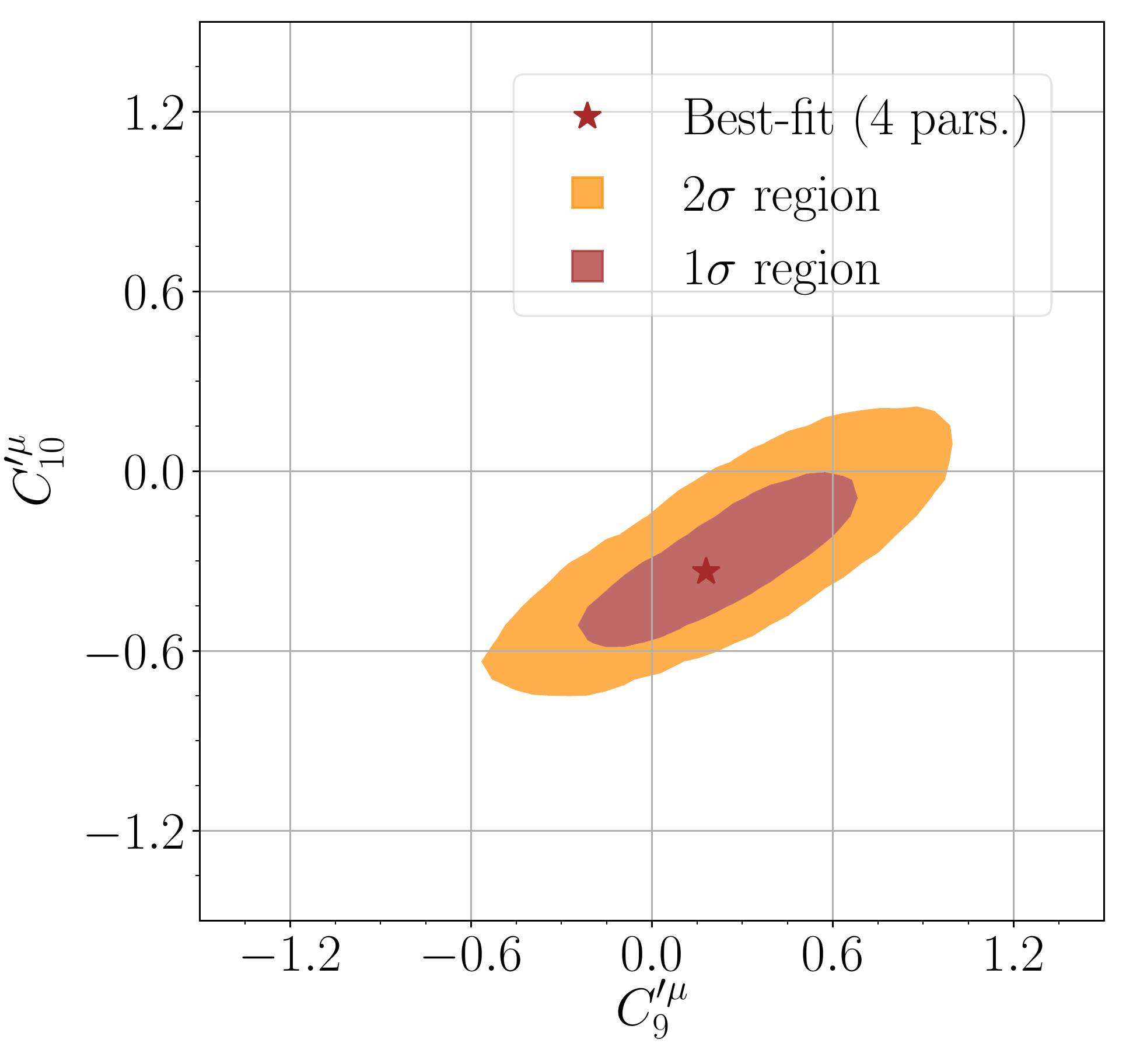}
}%
\caption{\footnotesize (a) In green, the $1\,\sigma$ (dark) and $2\,\sigma$ (light) credible regions of the posterior pdf for the scan in the input parameters $C_9^{\mu}$, $C_{10}^{\mu}$ 
(third row of \reftable{tab:priors}), 
compared with the marginalized 2-dimensional regions in the same parameters for 
the scan with $C_9^{\mu}$, $C_{10}^{\mu}$, $C_9^{\prime\mu}$, $C_{10}^{\prime\mu}$ all floating (fifth row of \reftable{tab:priors}), 
which are shown in brown ($1\,\sigma$) and orange ($2\,\sigma$).
The red stars mark the position of the best-fit points. 
(b) A similar comparison of the posterior pdf for the scan in $C_9^{\mu}$, $C_9^{\prime\mu}$ (shades of green) and the one 
with $C_9^{\mu}$, $C_{10}^{\mu}$, $C_9^{\prime\mu}$, $C_{10}^{\prime\mu}$ all floating (orange/brown). (c) 
The marginalized 2-dimensional credible regions in $C_9^{\prime\mu}$, $C_{10}^{\prime\mu}$ for 
the scan with $C_9^{\mu}$, $C_{10}^{\mu}$, $C_9^{\prime\mu}$, $C_{10}^{\prime\mu}$ all floating (fifth row of \reftable{tab:priors}).
\label{fig:4pars}}
\end{figure*}

This is presented in \reffig{fig:4pars}. In \reffig{fig:4pars}(a) we show a comparison between the marginalized pdf in the 
($C_9^{\mu}$, $C_{10}^{\mu}$) plane for the  scan with 2 input NP parameters (third row of \reftable{tab:priors}), 
and the one with 4 NP parameters (fifth row of \reftable{tab:priors}). 
Larger negative values of $C_9^{\mu}$ are favored by the data with 4 parameters. 
In \reffig{fig:4pars}(b) we show an equivalent comparison in the ($C_9^{\mu}$, $C_9^{\prime \mu}$) plane, between the marginalized posterior pdf of the 2-parameter scan 
(fourth row of \reftable{tab:priors}), versus the 4-parameter scan (fifth row of \reftable{tab:priors}). 
An ample region of the parameter space with $C_9^{\prime \mu}\leq 0$ appears, due to the introduction of the parameter $C_{10}^{\prime \mu}$.
The correlation with $C_9^{\prime \mu}$ is explicitly shown in \reffig{fig:4pars}(c)
and can be easily inferred from \refeq{RK_form}, as the product $C_9^{\prime\mu}-C_{10}^{\prime\mu}$ takes the role of the Wilson coefficient $C_9^{\prime\mu}$ in the 2-parameter scan.

\begin{figure*}[t]
\centering
\subfloat[]{%
\includegraphics[width=0.35\textwidth]{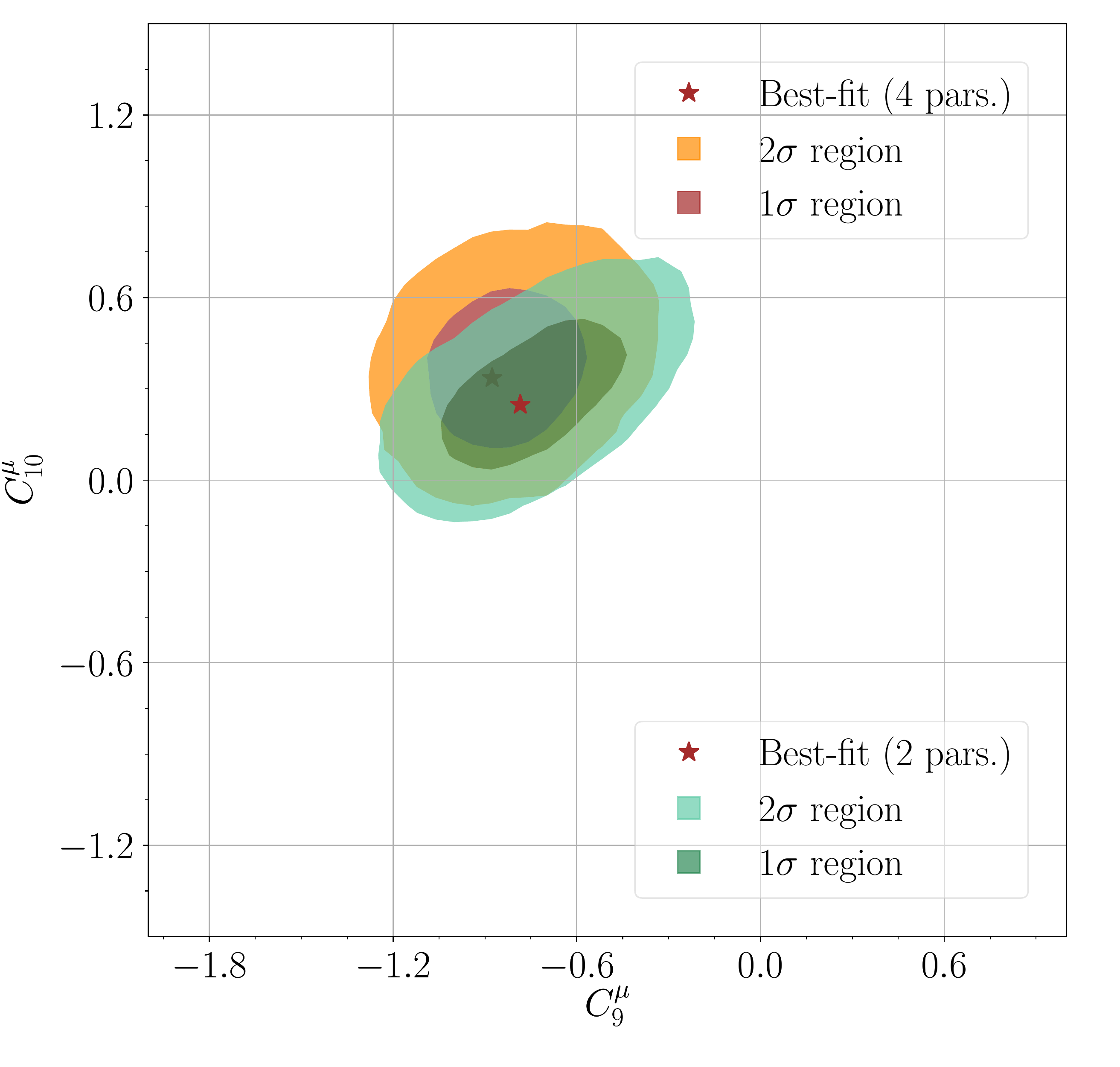}
}%
\hspace{0.02\textwidth}
\subfloat[]{%
\includegraphics[width=0.35\textwidth]{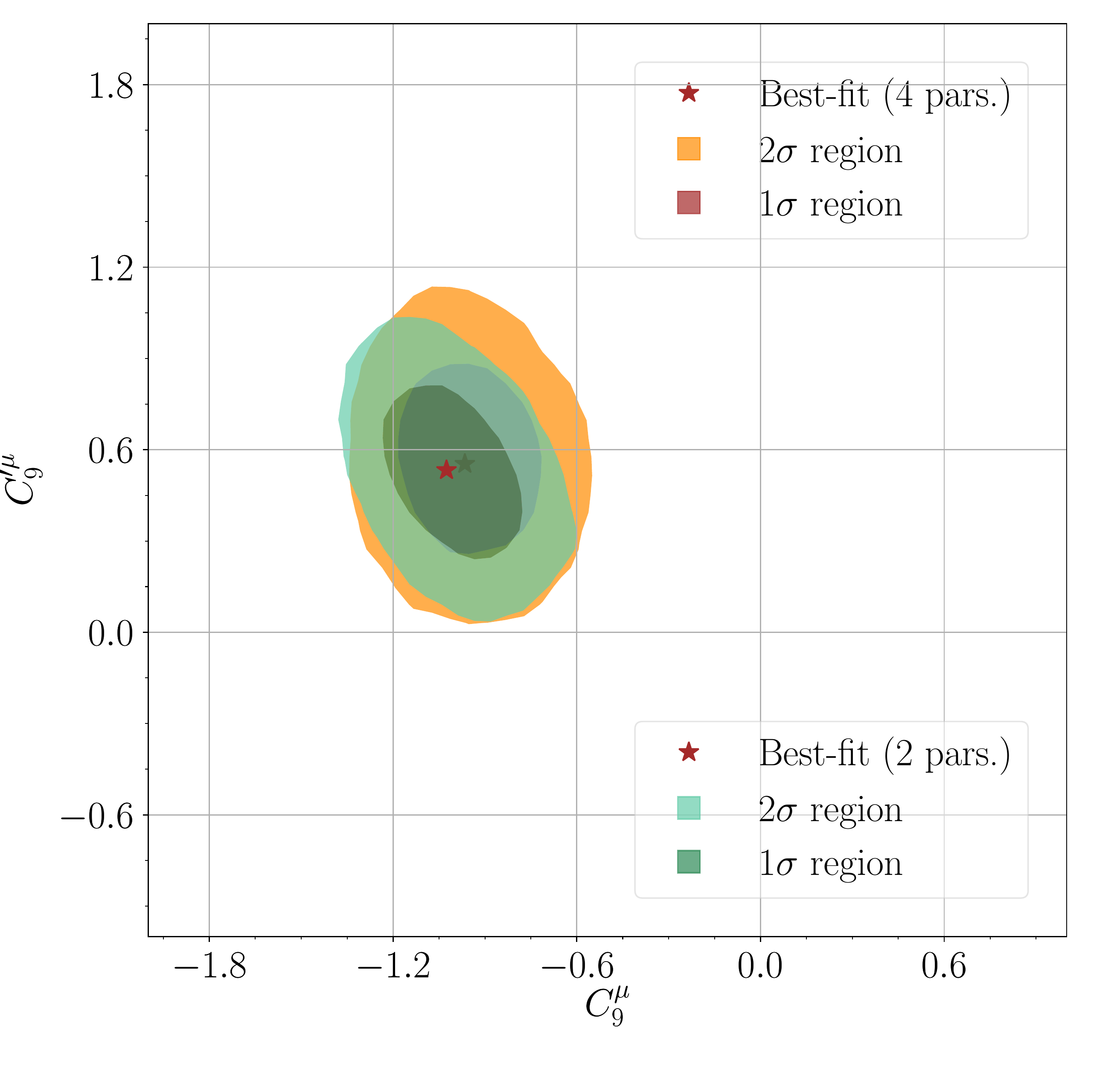}
}%
\\
\subfloat[]{%
\includegraphics[width=0.35\textwidth]{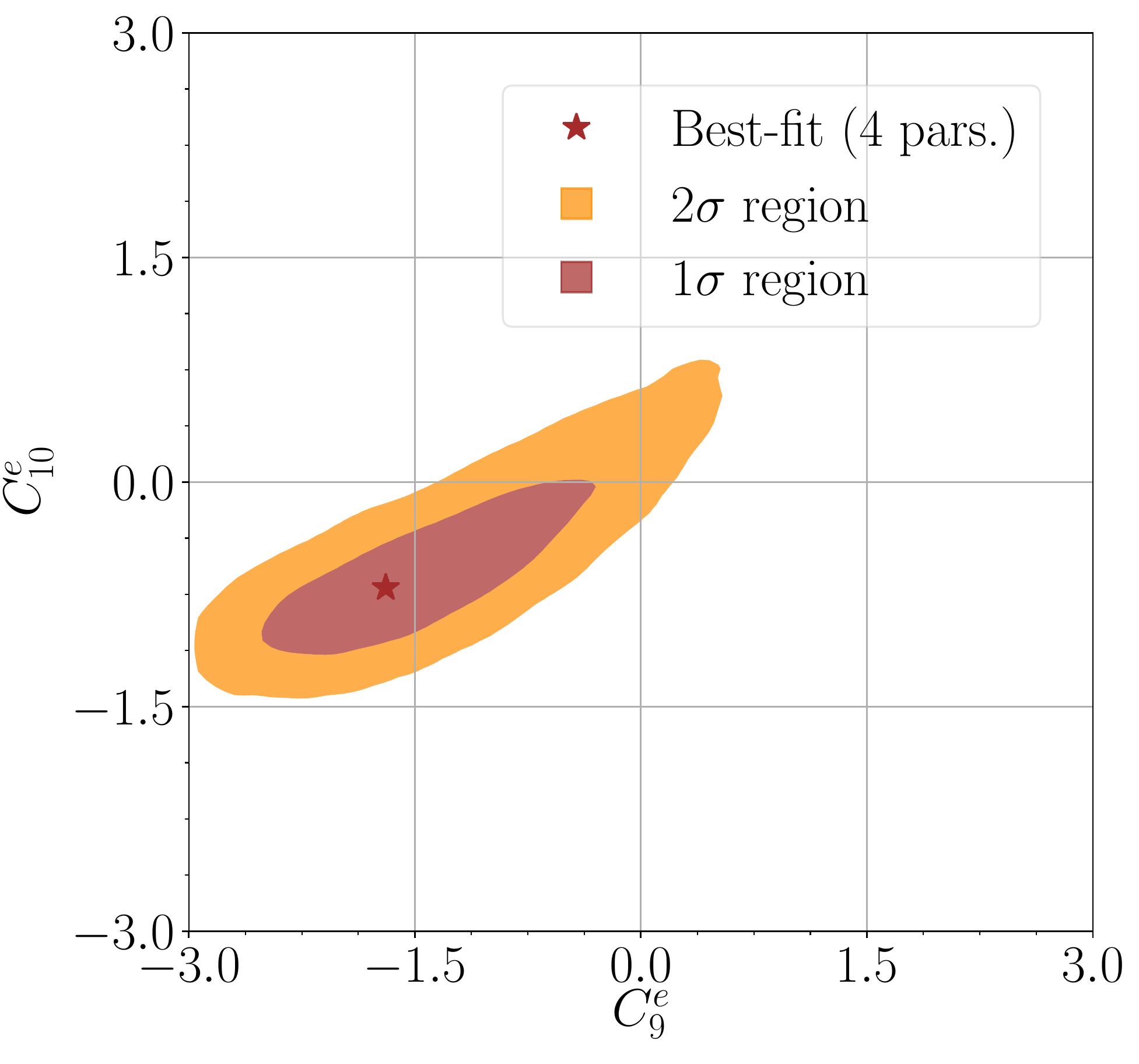}
}%
\hspace{0.02\textwidth}
\subfloat[]{%
\includegraphics[width=0.35\textwidth]{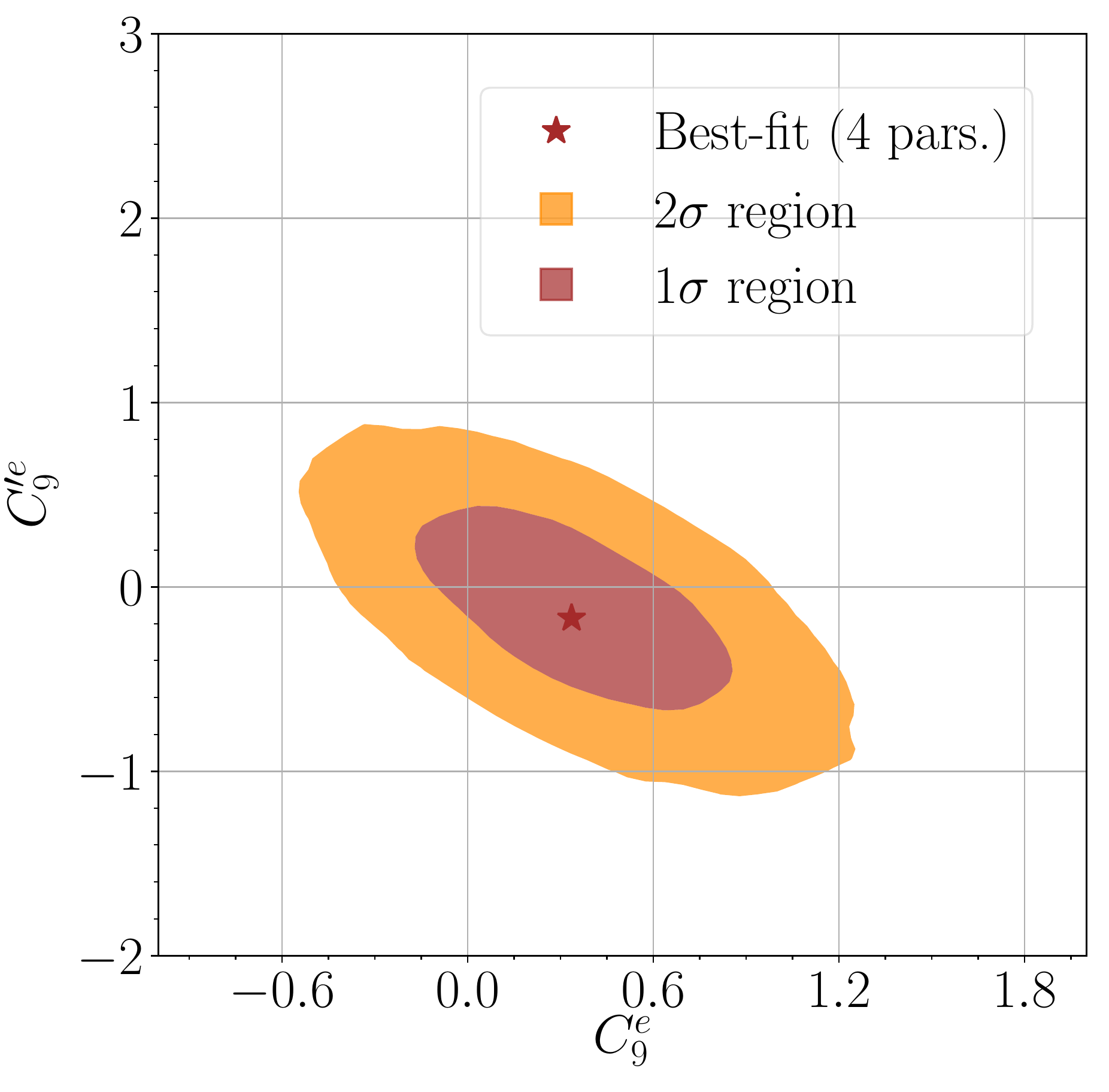}
}%
\caption{\footnotesize (a) In green, the $1\,\sigma$ (dark) and $2\,\sigma$ (light) credible regions of the posterior pdf for the scan in the input parameters $C_9^{\mu}$, $C_{10}^{\mu}$ 
(third row of \reftable{tab:priors}), 
compared with the marginalized 2-dimensional regions in the same parameters for 
the scan with $C_9^{\mu}$, $C_{10}^{\mu}$, $C_9^{e}$, $C_{10}^{e}$ all floating (sixth row of \reftable{tab:priors}), 
which are shown in brown ($1\,\sigma$) and orange ($2\,\sigma$).
The red stars mark the position of the best-fit points. 
(b) A similar comparison of the posterior pdf for the scan in $C_9^{\mu}$, $C_9^{\prime\mu}$ (fourth row of \reftable{tab:priors}, in shades of green) and the one 
with $C_9^{\mu}$, $C_{9}^{\prime\mu}$, $C_9^{e}$, $C_{9}^{\prime e}$ all floating (seventh row of \reftable{tab:priors}, in orange/brown). (c) 
The marginalized 2-dimensional credible regions in $C_9^{e}$, $C_{10}^{e}$ for 
the scan with $C_9^{\mu}$, $C_{10}^{\mu}$, $C_9^{e}$, $C_{10}^{e}$ all floating. (d) The marginalized 2-dimensional credible regions in $C_9^{e}$, $C_{9}^{\prime e}$ for 
the scan with $C_9^{\mu}$, $C_{9}^{\prime\mu}$, $C_9^{e}$, $C_{9}^{\prime e}$ all floating.
\label{fig:4parsb}}
\end{figure*}

The 2-dimensional regions of the posterior pdf undergo less dramatic modifications if one scans a different set of 
4 input parameters: $C_9^{\mu}$, $C_{10}^{\mu}$, $C_9^{e}$, $C_{10}^{e}$, see the sixth row of \reftable{tab:priors}, or 
$C_9^{\mu}$, $C_{9}^{\prime\mu}$, $C_9^{e}$, $C_{9}^{\prime e}$, see the seventh row of \reftable{tab:priors}. 
We perform the comparison between the relative marginalized 2-dimensional regions of these different models in 
\reffig{fig:4parsb}(a) and \reffig{fig:4parsb}(b). The details of the plots are explained in the caption.
Note, that the Wilson coefficients of the electron sector, whose pdf's are presented in \reffig{fig:4parsb}(c) and \reffig{fig:4parsb}(d) remain consistent at $2\,\sigma$
with zero, implying that the global data set can be easily explained by the presence of NP in the muon sector only.

Incidentally, it is important to point out that it is in principle possible to fit 
the discrepancies from the SM observed in $R_K$ and $R_{K^{\ast}}$ with the 4 Wilson coefficients of the electron sector alone\cite{Altmannshofer:2017yso,DAmico:2017mtc}. 
However, in doing so one encounters significant tension with the experimental determination of the observables 
tabularized in Tables~\ref{tab:data_B0_exp14}-\ref{tab:data_B0_exp13} of \ref{app:obs},
which do not present deviations from the expected SM value. For this reason 
the maximum of the likelhood function lies 
squarely in the regions where only the coefficients of the muon sector are nonzero.

\begin{figure*}[t]
\centering
\subfloat[]{%
\includegraphics[width=0.35\textwidth]{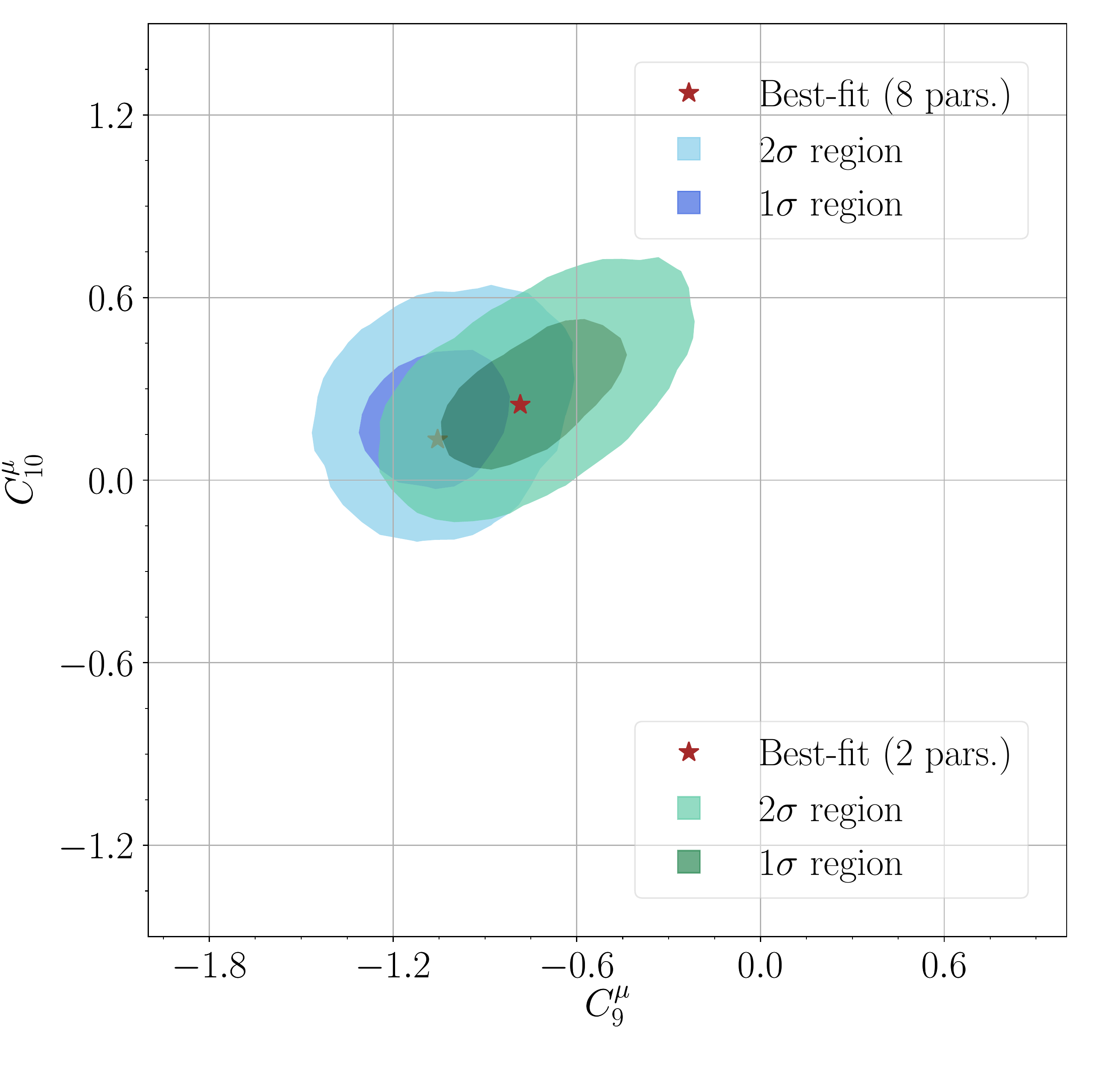}
}%
\hspace{0.02\textwidth}
\subfloat[]{%
\includegraphics[width=0.35\textwidth]{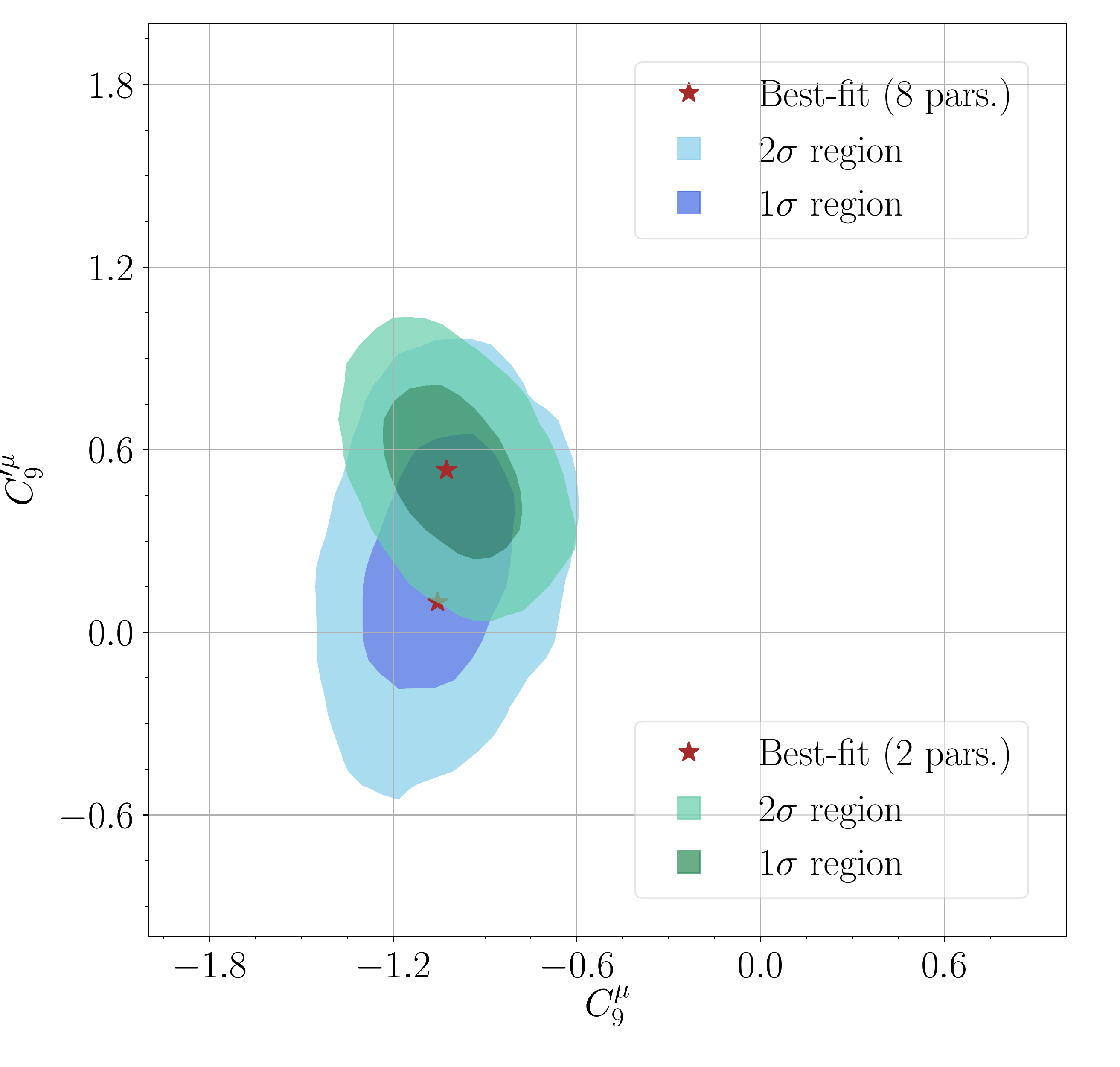}
}%
\caption{\footnotesize (a) In green, the $1\,\sigma$ (dark) and $2\,\sigma$ (light) credible regions of the posterior pdf for the scan in the input parameters $C_9^{\mu}$, $C_{10}^{\mu}$ 
(third row of \reftable{tab:priors}), 
compared with the marginalized 2-dimensional regions in the same parameters for 
the scan with all 8 NP parameters floating (eighth row of \reftable{tab:priors}), 
which are shown in shades of blue.
The stars mark the position of the best-fit points. 
(b) A similar comparison of the posterior pdf for the scan in $C_9^{\mu}$, $C_9^{\prime\mu}$ (fourth row of \reftable{tab:priors}, shades of green) and the one 
with 8 parameters floating (eighth row of \reftable{tab:priors}, shades of blue). 
\label{fig:8pars}}
\end{figure*}

We finally present in shades of blue the marginalized pdf of the 8-parameter scan introduced in the eighth row of \reftable{tab:priors} in the most relevant planes,  
($C_9^{\mu}$, $C_{10}^{\mu}$) in \reffig{fig:8pars}(a), and ($C_9^{\mu}$, $C_9^{\prime\mu}$) in \reffig{fig:8pars}(b). The posterior regions are 
compared to the $1\,\sigma$ and $2\,\sigma$ regions of the 2-parameters scans in the same plane, presented in shades of green. As one can see, the figures do not differ significantly 
from \reffig{fig:4pars}(a) and \reffig{fig:4pars}(b), as can be expected given the limited impact the Wilson coefficient of the electron sector bring to the fit.

\begin{table*}[t]
	\begin{center}
		\begin{tabular*}{0.64\textwidth}{c|c|ccc|ccccc}
         \toprule                     
         \bf Input parameters & $-$\textbf{ln} $\mathcal{Z}$ & \bf Pull & $\bm{\chi^2_{\textrm{TOT}}}$ & $\bm{\frac{\chi^2_{\textrm{TOT}}}{d.o.f}}$ &  \footnotesize  $\bm{\chi^2_{\mu}}$ & \footnotesize$\bm{\chi^2_{e}}$ & \footnotesize$\bm{\chi^2_{R_{K}}}$ & \footnotesize$\bm{\chi^2_{R_{K^*}}}$ \\
		\toprule
            \bf SM  &$88.5$ & $-$ &  $174.7$ & $1.29$ & $145.7$ & $6.5$ & $8.1$ & $12.0$ \\
            \rowcolor{LightGray}
            \cellcolor{White} & $88.3$ & $-$ & $174.4$ & $1.24$ &  $145.7$ & $6.5$ & $6.2$ & $13.6$\\
            \hline
            $\bm{C_9^{\mu}}$ &  $75.8$  & $5.0\,\sigma$ &$145.6$ &$1.09$ & $132.5$ & $6.7$ & $0.2$ & $6.0$\\
            \rowcolor{LightGray}
            \cellcolor{White} & $77.3$ &$4.7\,\sigma$ & $148.4$ & $1.06$ &  $132.2$ & $6.6$ & $0.3$ & $8.9$\\ 
            \hline
            $\bm{C_9^{\mu}=-C_{10}^{\mu}}$ &  $74.4$  & $5.3\,\sigma$ &$142.4$ &$1.06$ & $132.4$ & $6.8$ & $0.2$ & $3.0$\\
            \rowcolor{LightGray}
            \cellcolor{White} & $77.5$ &$4.8\,\sigma$ & $148.2$ & $1.06$ &  $133.2$ & $6.7$ & $1.2$ & $7.0$\\            
            \hline
            $\bm{C_9^{\mu},\,C_{10}^{\mu}}$ & $74.5$& $5.3\,\sigma$ & $140.1$ & $1.05$ & $129.8$ & $6.8$ & $0.2$& $3.4$ \\
            \rowcolor{LightGray}
            \cellcolor{White} & $77.6$ &$4.7\,\sigma$ & $146.1$ & $1.05$ &  $130.3$ & $6.7$ & $1.5$ & $7.6$\\
            \hline
            $\bm{C_9^{\mu},\,C_{9}^{\prime\mu}}$ & $75.1$ & $5.2\,\sigma$ & $141.1$ & $1.06$ & $128.1$ & $6.7$ & $2.0$& $4.1$ \\
            \rowcolor{LightGray}
            \cellcolor{White} &$75.8$ &$5.0\,\sigma$& $142.3$ & $1.02$ & $127.6$ & $6.7$ & $0.5$ & $7.3$ \\
            \hline
            $\bm{C_{9}^{\mu},\,C_{10}^{\mu},\,C_{9}^{\prime\mu},\,C_{10}^{\prime\mu}}$ & $74.0$ & $5.4\,\sigma$ &$133.3$ & $1.02$  &$123.5$ & $6.8$& $0.6$ & $2.4$ \\
            \rowcolor{LightGray}
            \cellcolor{White}
             &$76.0$ & $5.1\,\sigma$ & $136.8$ & $1.00$ &$123.2$ & $6.8$ & $0.0$ & $6.8$\\
            \hline
            $\bm{C_{9}^{\mu},\,C_{10}^{\mu},\,C_{9}^{e},\,C_{10}^{e}}$ & $75.6$ &  $4.9\,\sigma$ & $138.8$ & $1.06$ & $129.7$ & $6.9$ & $0.0$ & $2.1$ \\
            \rowcolor{LightGray}
            \cellcolor{White} & $78.0$& $4.5\,\sigma$&$142.7$ & $1.04$ &$129.8$ & $7.1$& $0.1$ & $5.8$\\
            \hline
            $\bm{C_{9}^{\mu},\,C_9^e,\,C_{9}^{\prime \mu},\,C_{9}^{\prime e}}$ & $75.8$ & $4.9\,\sigma$ & $138.5$ & $1.06$ & $127.5$ & $7.8$ & $0.5$ & $2.4$\\
            \rowcolor{LightGray}
            \cellcolor{White} & $77.7$& $4.6\,\sigma$ & $141.6$ &$1.03$ & $127.2$ & $7.0$& $0.2$& $6.7$\\
            \hline
            $\bm{C_{9}^{\mu},\,C_{10}^{\mu},\,C_{9}^{\prime\mu},\,C_{10}^{\prime\mu}}$ & $76.2$ &$4.7\,\sigma$ &$132.4$ &$1.04$ & $123.3$ & $6.7$ & $0.3$ & $2.1$ \\
            \rowcolor{LightGray}
            \cellcolor{White} 
            $\bm{C_{9}^{e},\,C_{10}^{e},\,C_{9}^{\prime e},\,C_{10}^{\prime e}}$ & $78.3$ &$4.4\,\sigma$ &$135.4$ & $1.02$& $123.3$& $6.6$& $0.2$ & $ 5.4$ \\
            \toprule
		\end{tabular*}
		\caption{Bayesian evidence, pull from the SM, and chi-squared statistics for the best-fit points of the considered scenarios. Gray highlighted rows correspond to the new data, while the white ones show the previous determinations.}
		\label{tab:results_chi2}
	\end{center}
\end{table*}

\begin{table*}[t]
	\begin{center}
		\begin{tabular*}{0.88\textwidth}{c|cccccccc|ccc} 
		    \toprule
            \bf{Input parameters} & $\bm C_9^{\mu}$ & $\bm C_{10}^{\mu}$ & $\bm C_9^{\prime\mu}$ & $\bm C_{10}^{\prime\mu}$ & $\bm C_9^{e}$ & $\bm C_{10}^{e}$ & $\bm C_9^{\prime e}$ & $\bm C_{10}^{\prime e}$ & \footnotesize$\bm R_{K}^{[1.1,6]}$ &\footnotesize $\bm R_{K*}^{low}$ & \footnotesize$\bm R_{K*}^{[1.1,6]}$ \\
		    \toprule
		    $\bm{C_9^{\mu}}$ & $-1.02$ & $0$ & $0$ & $0$ & $0$ & $0$  & $0$  & $0$ &$0.789$ & $0.893$ & $0.844$\\
            \rowcolor{LightGray}
            \cellcolor{White} & $-0.90$ & $0$ & $0$ & $0$ & $0$ & $0$  & $0$  & $0$  & $0.811$ & $0.896$& $0.858$\\
            \hline
		    $\bm{C_9^{\mu}=-C_{10}^{\mu}}$ & $-0.64$ & $0.64$ & $0$ & $0$ & $0$ & $0$  & $0$  & $0$ &$0.710$ & $0.852$ & $0.720$\\
            \rowcolor{LightGray}
            \cellcolor{White} & $-0.48$ & $0.48$ & $0$ & $0$ & $0$ & $0$  & $0$  & $0$  & $0.777$ & $0.869$& $0.783$\\
            \hline
            $\bm{C_9^{\mu},\,C_{10}^{\mu}}$ & $-0.91$ & $0.42$ & $0$ & $0$ & $0$ & $0$  & $0$  & $0$ &$0.707$ & $0.862$ & $0.740$\\
            \rowcolor{LightGray}
            \cellcolor{White} & $-0.78$ & $0.25$ & $0$ & $0$ & $0$ & $0$  & $0$  & $0$  & $0.771$ & $0.878$& $0.802$\\
            \hline
            $\bm{C_9^{\mu},\,C_{9}^{\prime\mu}}$ & $-1.08$ & $0$ & $0.49$ & $0$ & $0$ & $0$  & $0$  & $0$ & $0.873$ & $0.870$ & $0.781$\\
            \rowcolor{LightGray}
            \cellcolor{White} & $-1.03$ & $0$& $0.53$& $0$& $0$& $0$&$0$ &$0$ & $0.891$ &$0.869$& $0.782$\\
            \hline
            $\bm{C_{9}^{\mu},\,C_{10}^{\mu},\,C_{9}^{\prime\mu},\,C_{10}^{\prime\mu}}$ & $-1.14$& $0.28$ &$0.21$ &$-0.31$ & $0$& $0$& $0$& $0$ & $0.812$&$0.835$ & $0.666$\\
            \rowcolor{LightGray}
            \cellcolor{White} & $-1.06$ & $0.18$ & $0.18$ &$-0.34$ & $0$& $0$& $0$& $0$& $0.855$& $0.844$ & $0.699$\\
            \hline
            $\bm{C_{9}^{\mu},\,C_{10}^{\mu},\,C_{9}^{e},\,C_{10}^{e}}$ & $-0.92$ & $0.40$ & $0$ & $0$ & $-1.50$& $-0.90$ & $0$ & $0$ & $0.733$ &$ 0.825$ & $0.680$\\
             \rowcolor{LightGray}
            \cellcolor{White} & $-0.88$& $0.34$& $0$& $0$& $-1.69$& $-0.71$& $0$&$0$ & $ 0.831$ &$0.848$& $0.756$\\
            \hline
            $\bm{C_{9}^{\mu},\,C_9^e,\,C_{9}^{\prime \mu},\,C_{9}^{\prime e}}$ & $-1.02$ & $0$ & $0.54$ & $0$ & $0.58$& $0$ & $-0.17$ & $0$ & $0.811$&$0.833$ &  $0.675$\\
             \rowcolor{LightGray}
            \cellcolor{White} &$-0.97$ & $0$&$0.55$ &$0$ & $0.34$&$0$ & $-0.17$& $0$& $0.873$ & $0.844$ & $0.712$\\
            \hline
            $\bm{C_{9}^{\mu},\,C_{10}^{\mu},\,C_{9}^{\prime\mu},\,C_{10}^{\prime\mu}}$ &$-1.10$ & $0.21$&  $0.21$ & $-0.30$& $-0.80$& $-0.63$ & $-0.73$& $-0.57$ & $0.792$ &$0.819$ & $0.646$\\
             \rowcolor{LightGray}
            \cellcolor{White}
            $\bm{C_{9}^{e},\,C_{10}^{e},\,C_{9}^{\prime e},\,C_{10}^{\prime e}}$ & $-1.05$&$0.13$& $0.10$ &$-0.38$& $-2.18$ & $-0.07$&$-2.73$& $-1.34$ & $0.875$ &$0.826$& $0.700$\\
            \hline
		\end{tabular*}
		\caption{Wilson coefficients at the best-fit points, as well as the values there of $R_K$ and $R_{K^{\ast}}$. Gray highlighted rows correspond to the new data, while the white ones show the previous determinations.}
		\label{tab:results_BF} 
	\end{center}
\end{table*}

We summarize the main characteristics of the 8 fits analyzed in this section in \reftable{tab:results_chi2}. 
The main Bayesian quantity 
is the negative logarithm of the evidence $\mathcal{Z}$, featured in the second column. We use Jeffrey's scale\cite{Jeffery:1998,Kass}
to interpret the Bayes factor, defined in \refsec{sec:method}, which will point to which model is favored by the data. 
In general, models that are characterized by a smaller number of input parameters 
tend to fare significantly better than those with a larger input set, as the latter are penalized by volume effects. Unless, of course, these volume effects are counterbalanced 
by a significantly higher likelihood function.   

In the specific cases considered here we see that, for example, 
\bea
\frac{p(d)_{C_9^{\mu},C_{9}^{\prime\mu}}}{p(d)_{C_9^{\mu},C_{10}^{\mu}}}\equiv\frac{\mathcal{Z}_{C_9^{\mu},C_{9}^{\prime\mu}}}{\mathcal{Z}_{C_9^{\mu},C_{10}^{\mu}}}&=&6.0 \quad (\textrm{Substantial})\nonumber\\
\frac{\mathcal{Z}_{C_9^{\mu},C_{10}^{\mu},C_9^{\prime\mu},C_{10}^{\prime\mu}}}{\mathcal{Z}_{C_9^{\mu},C_{10}^{\mu}}}&=&5.0 \quad (\textrm{Substantial})\nonumber\\
\frac{\mathcal{Z}_{C_9^{\mu},C_{9}^{\prime\mu}}}{\mathcal{Z}_{C_9^{\mu},C_{10}^{\mu},C_9^{\prime\mu},C_{10}^{\prime\mu}}}&=&1.2 \quad (\textrm{Barely worth mentioning})\nonumber\\
\frac{\mathcal{Z}_{C_9^{\mu},C_{10}^{\mu},C_9^{\prime\mu},C_{10}^{\prime\mu}}}{\mathcal{Z}_{C_9^{\mu},C_{10}^{\mu},C_9^{e},C_{10}^{e}}}&=&7.4 \quad (\textrm{Substantial})\nonumber\\
\frac{\mathcal{Z}_{C_9^{\mu},C_{10}^{\mu},C_9^{\prime\mu},C_{10}^{\prime\mu}}}{\mathcal{Z}_{C_9^{\mu},C_{9}^{\prime\mu},C_9^{e},C_{9}^{\prime e}}}&=&5.6\,, \quad (\textrm{Substantial})\label{BayesComp}
\eea
where in parentheses we report the tabularized ``strength of evidence'' according to Jeffrey's scale. From Eqs.~(\ref{BayesComp}) 
one draws the conclusion that the models slightly favored by the data are the ones in the fifth and sixth row of \reftable{tab:results_chi2}.

We note here that the new data continues to favor strongly NP scenarios over to the SM alone. The Bayes factor over the SM for the least likely NP case (8-parameter scan, ninth row of \reftable{tab:results_chi2}) exceeds $10^4:1$. This reads ``decisive'' on the tabularized Jeffrey's scale.  In the case of the most likely scenarios of \reftable{tab:results_chi2}
the Bayes factor over the SM increases further, by more than one order of magnitude.
We also note in the first of relations~(\ref{BayesComp}), that after the data upgrade the 2-parameter NP scenario in $C_9^{\mu},C_9'^{\mu}$ has become favored over the 2-parameter model in $C_9^{\mu},C_{10}^{\mu}$.
The reason can be easily inferered, again, from Eqs.~(\ref{RK_form})-(\ref{RK_form1}): since $R_K$ and $R_{K^{\ast}}$ are expected to be equal in the $C_9^{\mu},C_{10}^{\mu}$ model and 
instead the new measurement of $R_K$ shows a slight shift towards the SM value, 
this scenario receives a penalty with respect to the previous determination.
This is not the case for the model with $C_9^{\mu},C_9'^{\mu}$, where $R_K$ and $R_{K^{\ast}}$ can be fitted individually with an appropriate choice of the input parameters. 
It will be interesting to see if the same trend is confirmed by the Run~2 determination of $R_{K^{\ast}}$ at LHCb.

In the remaining columns of \reftable{tab:results_chi2} we make contact with frequentist approaches by presenting the pull of the best-fit point from the SM, 
the minimum chi-squared $\chisq_{\textrm{TOT}}$, 
the minimum chi-squared per degree of freedom,
and the relative chi-squared of the muon observables, $\chisq_{\mu}$, electron observables, $\chisq_{e}$, and LFUV observables, $\chisq_{R_K}$ and $\chisq_{R_K^{\ast}}$, of the 9
scans analyzed here.
We calculate the minimum chi-squared per degree of freedom very roughly, neglecting all correlations, 
as an indicative measure of the relative goodness of fit: 
\be
\frac{\chisq_{\textrm{TOT}}}{\textrm{d.o.f.}}=\frac{\chisq_{\textrm{TOT}}}{\textrm{num.constraints + 1 - (num.input + 1)}}\,,
\ee
where the $\pm1$ is placed as a reminder of the nuisance parameter. The full list of constraints is collected in \ref{app:obs}.

Finally, also to favor the comparison with frequentist analyses, 
we show in \reftable{tab:results_BF} the numerical value of the Wilson coefficients and of the most important observables 
at the best-fit points.

For the scans with 1 or 2 independent Wilson coefficients, our results are in good agreement with those reported in Refs.\cite{Alguero:2019ptt,Alok:2019ufo,Ciuchini:2019usw,Datta:2019zca,Aebischer:2019mlg}. In general, we observe slightly lower best-fit point values of the coefficient $C_9^{\mu}$. This is due to the fact that in our analysis we consider a floating nuisance parameter $V_{cb}$, whose lower values facilitate the fitting of the experimental data. For example, for the scenario  in the first row of \reftable{tab:priors}, the best-fit $V_{cb}$ is $0.5\,\sigma$ away from its central value. Note also that, for the same reason, the NP pull from the SM is in our case systematically lower than in Refs.\cite{Alguero:2019ptt,Datta:2019zca,Aebischer:2019mlg}, as the presence of an additional input parameter allows to improve the benchmark fit of the SM. 

\section{Simple models for $b \to s ll$ anomalies\label{sec:mod}}

In the previous section we determined the preferred $1\,\sigma$ and $2\,\sigma$ ranges for the NP Wilson coefficients relevant to explaining $b\to s $ anomalies. In the following we will discuss several simple BSM 
scenarios that are known to naturally lead to the desired EFT operator structure due to an exchange of a BSM boson with flavor-violating couplings to the SM quarks and leptons. 
Since we confirmed in our model-independent fit (see \reftable{tab:results_chi2}) that the presence of the electron Wilson coefficients does not improve the goodness of the fit, 
we will only consider models in which the BSM boson couples exclusively to muons.

\subsection{Heavy $Z'$}

As a first example, we discuss the SM extended by an additional vector boson, commonly denoted as $Z'$.
The most generic Lagrangian, parametrizing LFUV couplings of $Z'$ to the $b$-$s$ current and the muons reads,
\begin{multline}
\mathcal{L} \supset  Z'_{\alpha}\left(\Delta_L^{sb}\, \bar{s}_L\gamma^{\alpha}\,b_L+ \Delta_R^{sb}\, \bar{s}_R\gamma^{\alpha}\,b_R+\textrm{H.c.}\right)\\
+ Z'_{\alpha}\left(\Delta^{\mu\mu}_{L}\,\bar{\mu}_L\gamma^{\alpha}\mu_L+\Delta^{\mu\mu}_{R}\,\bar{\mu}_R\gamma^{\alpha}\mu_R\right)\,.
\end{multline}

The relevant NP Wilson coefficients are then given by
\bea\label{C9generic}
C_{9}^{\mu}&=&-2\frac{\Delta_{L}^{sb}\Delta^{\mu\mu}_9}{V_{tb}V_{ts}^{\ast}}\left(\frac{\Lambda_v}{m_{Z'}}\right)^2, \\
C_{9}^{\prime \mu} &=&-2\frac{\Delta_{R}^{sb}\Delta^{\mu\mu}_9}{V_{tb}V_{ts}^{\ast}}\left(\frac{\Lambda_v}{m_{Z'}}\right)^2 \, ,\\
C_{10}^{\mu}&=&-2\frac{\Delta_{L}^{sb}\Delta^{\mu\mu}_{10}}{V_{tb}V_{ts}^{\ast}}\left(\frac{\Lambda_v}{m_{Z'}}\right)^2, \\
C_{10}^{\prime \mu} &=&-2\frac{\Delta_{R}^{sb}\Delta^{\mu\mu}_{10}}{V_{tb}V_{ts}^{\ast}}\left(\frac{\Lambda_v}{m_{Z'}}\right)^2\,,
\eea
where $\Delta_9^{\mu\mu}\equiv (\Delta^{\mu\mu}_{R}+\Delta^{\mu\mu}_{L})/2$, $\Delta_{10}^{\mu\mu}\equiv (\Delta^{\mu\mu}_{R}-\Delta^{\mu\mu}_{L})/2$, $m_{Z'}$ 
is the mass of the $Z'$ boson, and
\be
\Lambda_v=\left(\frac{\pi}{\sqrt{2}G_F\alpha_{\textrm{em}}}\right)^{1/2}\approx 4.94\tev,
\ee
is the typical effective scale of the new physics.

If the heavy $Z'$ is the gauge boson of a new U(1)$_X$ gauge group, its couplings to the gauge eigenstates must be flavor-conserving, and 
an additional structure is required to generate $\Delta_{L}^{sb}$ and $\Delta_{R}^{sb}$. Thus, in this work we also 
consider the impact of the new LHCb and Belle data 
on the masses and couplings of a few simplified but UV complete models.

\subsubsection{Variations of the $L_{\mu}-L_{\tau}$ model}

\textbf{Model~1.} A U(1)$_X$ model that has proven to be quite popular is the 
traditional $X=L_{\mu}-L_{\tau}$ model\cite{Foot:1990mn,He:1990pn,He:1991qd,Altmannshofer:2014cfa}, in which the SM leptons carry an additional charge and 
are characterized by the following 
SU(3)$\times$SU(2)$_L\times$U(1)$_Y\times$U(1)$_X$ quantum numbers:
\bea
l_1:(\mathbf{1},\mathbf{2},-1/2,0) & \quad & e_R:(\mathbf{1},\mathbf{1},1,0)\label{electrons}\nonumber\\
l_2:(\mathbf{1},\mathbf{2},-1/2,1) & \quad & \mu_R:(\mathbf{1},\mathbf{1},1,-1)\label{muons}\nonumber\\
l_3:(\mathbf{1},\mathbf{2},-1/2,-1) & \quad & \tau_R:(\mathbf{1},\mathbf{1},1,1)\label{taus}\label{LMLT3}\,.
\eea
In the above and the following text we label SM fields by lower-case letters and BSM matter with capital ones.
Besides $Z'$, we also add to the SM a scalar singlet field $S$ to spontaneously break the U(1)$_X$ symmetry and 
VL quark pairs $Q,Q'$ and $D,D'$ to create the flavor-changing couplings $\Delta^{bs}_{L,R}$\cite{Fox:2011qd,Bobeth:2016llm}:
\bea
S:(\mathbf{1},\mathbf{1},0,-1)\,, & & \\
Q:(\mathbf{3},\mathbf{2},1/6,-1)\,, & \quad & Q':(\mathbf{\bar{3}},\mathbf{2},-1/6,1)\,, \\
D:(\mathbf{\bar{3}},\mathbf{1},1/3,-1)\,, & \quad & D':(\mathbf{3},\mathbf{1},-1/3,1)\,.\label{LMLTD}
\eea

Given the quantum numbers introduced in Eqs.~(\ref{LMLT3})-(\ref{LMLTD}),
the Lagrangian features new Yukawa couplings $\lam_{Q,i}$ and $\lam_{D,i}$ that mix the SM and BSM fields, as well as VL mass terms $M_{Q,D}$:
\be\label{LagrLMLT}
\mathcal{L}\supset -\lam_{Q,i}S Q' q_i-\lam_{D,i}S D' d_{R,i}-M_{Q} Q'Q-M_D D'D+\textrm{H.c.}\,,
\ee
where in writing down \refeq{LagrLMLT} we have adopted the Weyl 2-component spinor notation and all spinors are left-chiral.
The $q_i$ are SM SU(2) doublets, $d_{R,i}$ are SU(2) singlets, and $i=1,2,3$ label the SM generations. 

After spontaneously breaking U(1)$_X$ by a new scalar vacuum expectation value (vev) $v_S$, rotating the Lagrangian~(\ref{LagrLMLT}) to the mass basis, 
and retaining only the Yukawa couplings that yield $b\to s\mu\mu$ transitions,
one obtains flavor-generating couplings of the form
\bea
\Delta_L^{sb}&\approx &-g_X \frac{\lam_{Q,2}\lam_{Q,3}v_S^2}{2M_Q^2+\left(\lam_{Q,2}^2+\lam_{Q,3}^2\right)v_S^2}\,,\\
\Delta_R^{sb}&\approx &g_X \frac{\lam_{D,2}\lam_{D,3}v_S^2}{2M_D^2+\left(\lam_{D,2}^2+\lam_{D,3}^2\right)v_S^2}\,,
\eea
and
\be
\Delta_9^{\mu\mu}=g_X\quad\quad\quad \Delta_{10}^{\mu\mu}=0\,,
\ee
where $g_X$ is the gauge coupling of the U(1)$_X$ group.

By recalling that $v_S\equiv m_{Z'}/g_X$ one finally obtains
\bea
C_{9}^{\mu}&=&\frac{2\Lambda_v^2}{V_{tb}V_{ts}^{\ast}}\frac{\lam_{Q,2}\lam_{Q,3}}{2 M_Q^2+\left(\lam_{Q,2}^2+\lam_{Q,3}^2\right)v_S^2}\,,  \\
C_{9}^{\prime \mu}& =&-\frac{2\Lambda_v^2}{V_{tb}V_{ts}^{\ast}}\frac{\lam_{D,2}\lam_{D,3}}{2 M_D^2+\left(\lam_{D,2}^2+\lam_{D,3}^2\right)v_S^2}\,,
\eea
while $C_{10}^{\mu}=C_{10}^{\prime \mu}=0$.
Without loss of generality one can assume that the couplings of the second and third generations are unified, denoted as $\lam_{Q,D}$. 
Therefore, the model can be parametrized in terms of only 3 free parameters: 
$m_{Z'}/g_X$, $M_Q/\lam_Q$, and $M_D/\lam_D$. The scanning ranges imposed in the global fit are shown in~\reftable{tab:priors}.
We scan $m_{Z'}/g_X$ for values not smaller than 500\gev, to evade the strong bound from neutrino trident production\cite{Altmannshofer:2014pba}.

Recall, finally, that any scenario with a non-universal $\Delta^{bs}_{L,R}$ coupling is subject to the strong $2\sigma$ constraint from $B_s$ mixing\cite{Altmannshofer:2014rta,DiLuzio:2017fdq}:
$R_{BB}\leq 0.014$. In our VL model the latter can be expressed in terms of the Wilson coefficients as\cite{Buras:2012jb}:  
\begin{multline}
R_{BB}=\left(\frac{g_2^2\, S_0}{16 \pi^2}\right)^{-1}\frac{v_h^2 v_S^2}{4 \Lambda_v^4}\left[\left(C^{\mu}_{9,\textrm{NP}}\right)^2+\left(C'^{\mu}_{9,\textrm{NP}}\right)^2\right.\\
\left.+\,0.094\,C^{\mu}_{9,\textrm{NP}} C'^{\mu}_{9,\textrm{NP}}\right]\,,
\end{multline}
where $v_h$ is the Higgs vev and $S_0\approx 2.3$ is a loop factor. 
We impose an upper bound on the prior range of  $m_{Z'}/g_X$ at 5\tev, which, as will appear below, is large enough to encompass the $2\,\sigma$ region of the posterior pdf in its entirety
once we include the constraint 
from $B_s$ mixing into our likelihood function as a gaussian bound. 
\bigskip

\textbf{Model~2.} Another realization of the $L_{\mu}-L_{\tau}$ model we consider is 
an extension of the SM that -- besides featuring one pair of VL quark doublets, $Q,Q'$, 
to generate the flavor-violating coupling of the $Z'$, $\Delta^{bs}_L$ -- is characterized by 
one pair of VL U(1)$_X$ neutral leptons $E,E'$\cite{Altmannshofer:2016oaq,Darme:2018hqg}, 
which have to be SU(2) singlets for reasons that will be clear below. One has
\bea
S:(\mathbf{1},\mathbf{1},0,-1)\,,& & \\
Q:(\mathbf{3},\mathbf{2},1/6,-1)\,, & \quad & Q':(\mathbf{\bar{3}},\mathbf{2},-1/6,1)\,,\\
E:(\mathbf{1},\mathbf{1},1,0)\,, & \quad & E':(\mathbf{1},\mathbf{1},-1,0)\,.
\eea

The gauge-invariant Lagrangian terms involving these new leptons read, in the Weyl notation,
\be\label{Lagrlept}
\mathcal{L}\supset -\lam_{E,2}S^{\ast}E'\mu_R -\lam_{E,3}S E'\tau_R -\widetilde{Y}_E\phi^{\dag}l_1 E-M_E E'E+\textrm{H.c.}\,,
\ee
where $\phi$ is the Higgs doublet.

For the purpose of explaing the muon anomalies only the second-generation Yukawa couplings $\lam_{E,2}$ can be retained in \refeq{Lagrlept}: 
\bea
C_{9}^{\mu}&=&\frac{\Lambda_v^2}{V_{tb}V_{ts}^{\ast}}\left(\frac{\lam_{Q,2}\lam_{Q,3}}{2 M_Q^2+\left(\lam_{Q,2}^2+\lam_{Q,3}^2\right)v_S^2}\right) \nonumber\\
&&\times \left(1+\frac{2 M_E^2}{2 M_E^2+\lam_{E,2}^2 v_S^2}\right),\\
C_{10}^{\mu}&=&\frac{\Lambda_v^2}{V_{tb}V_{ts}^{\ast}}\left(\frac{\lam_{Q,2}\lam_{Q,3}}{2 M_Q^2+\left(\lam_{Q,2}^2+\lam_{Q,3}^2\right)v_S^2}\right)\nonumber\\
&&\times \left(-1+\frac{2 M_E^2}{2 M_E^2+\lam_{E,2}^2 v_S^2}\right).
\eea
Note that if we had chosen a VL lepton SU(2) doublet instead of a singlet we would have obtained $C_{9}^{\mu}$ and $C_{10}^{\mu}$ of the same sign,
which is disfavored by the data.

Again we parametrize this model in terms of 3 free parameters: 
$m_{Z'}/g_X$, $M_Q/\lam_Q$, and $M_E/\lam_{E,2}$, where $\lam_{Q}$ represent equal couplings to the second and third generation quarks. 
Their scanning ranges imposed in the global fit are shown in~\reftable{tab:priors}.
Again we include the bound from $B_s$ mixing in the likelihood function. 

\begin{figure*}[h]
\centering
\subfloat[]{%
\includegraphics[width=0.35\textwidth]{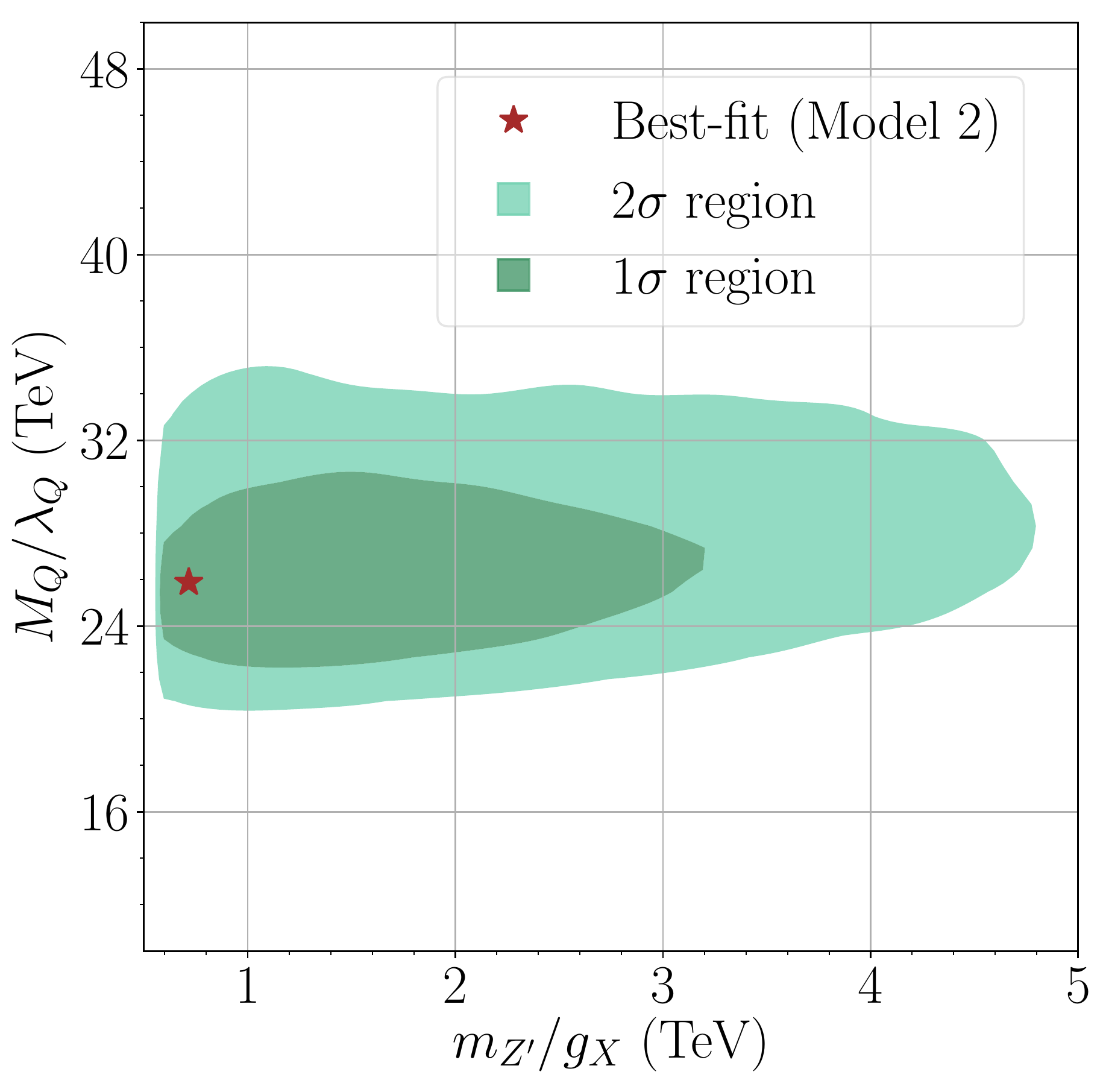}
}%
\\
\subfloat[]{%
\includegraphics[width=0.35\textwidth]{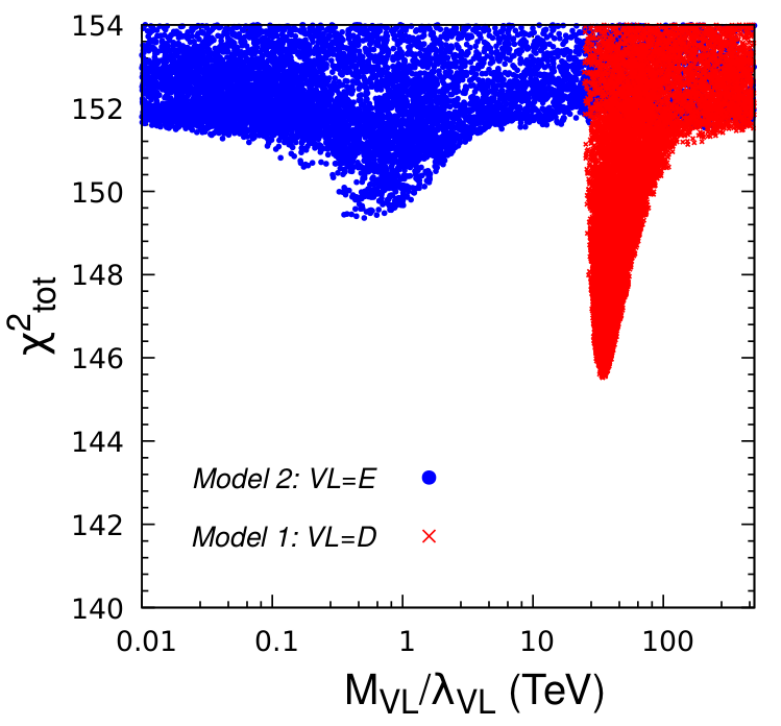}
}%
\hspace{0.02\textwidth}
\subfloat[]{%
\includegraphics[width=0.31\textwidth]{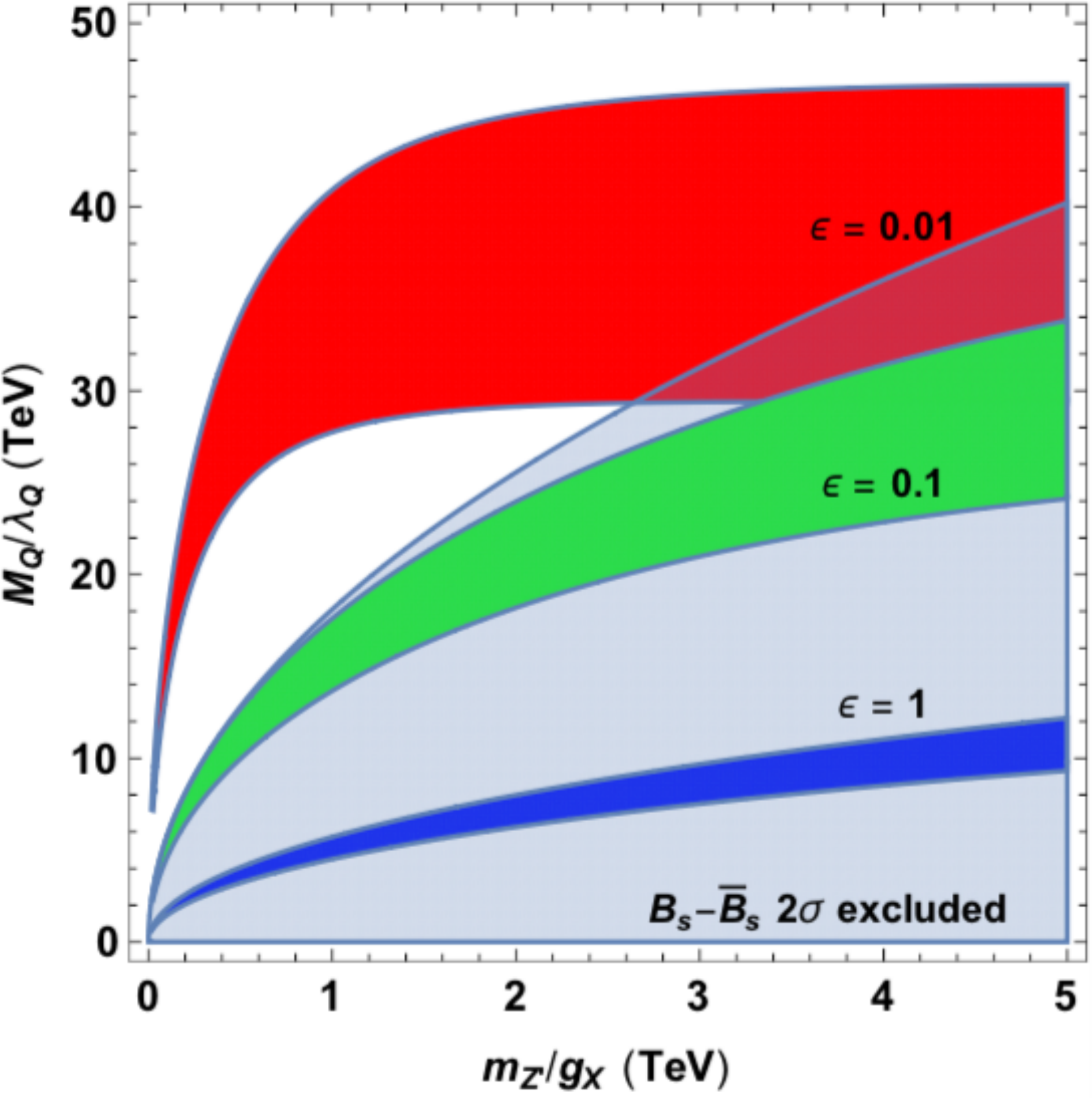}
}%
\caption{\footnotesize (a) The marginalized 2-dimensional posterior pdf in the ($m_{Z'}/g_X$, $M_Q/\lam_Q$) plane in Model~2 (the pdf is very similar in Model~1). 
(b) A scatter plot of the $\chisq$ distribution as a function of VL mass
rescaled by the Yukawa coupling for Model~1 (red, $M_D/\lam_D$ on the $x$-axis) and Model~2 (blue, $M_{E}/\lam_{E,2}$ on the $x$-axis). (c) The $2\,\sigma$ regions of the profile likelihood
in the ($m_{Z'}/g_X$, $M_Q/\lam_Q$) plane of Model~3 for different values of $\epsilon$ ($M_L/\lam_{L,2}=\epsilon M_Q/\lam_Q$). The gray area is excluded by the upper 
bound on $R_{BB}$ from $B_s$ mixing. 
\label{fig:mod}}
\end{figure*}

We present in \reffig{fig:mod}(a) the marginalized 2-dimensional posterior pdf in the ($m_{Z'}/g_X$, $M_Q/\lam_Q$) plane for Model~2. 
The VL mass range is determined by the $2\,\sigma$ range in $C_{9,\textrm{NP}}^{\mu}$ and lies around a 20--30\tev\ scale for a coupling $\lam_Q$ of order unity.
Within $2\,\sigma$ probability, the $m_{Z'}/g_X$ mass is limited to values below 5\tev, as a result of the $B_s$ mixing constraint, which directly depends on $v_S^2$.
Note that the  2-dimensional posterior pdf in the ($m_{Z'}/g_X$, $M_Q/\lam_Q$) plane for Model~1 looks very similar to \reffig{fig:mod}(a), as a direct consequence of the bounds
on $C_{9,\textrm{NP}}^{\mu}$ and $R_{BB}$.

In both Model~1 and Model~2, the second VL mass is unbounded from above at the $2\,\sigma$ level, 
as depicted in \reffig{fig:mod}(b), where we show scatter plots of the \chisq\ distributions of Model~1 (red points) and Model~2 (blue points) versus the VL mass rescaled by the Yukawa coupling.
This is a consequence of the fact that $C_{9,\textrm{NP}}^{\prime \mu}$ in Model~1 and, especially $C_{10,\textrm{NP}}^{\mu}$ in Model~2, are consistent with zero at the $2\,\sigma$ level. 
On the other hand, the values of $M_D/\lam_D$ and $M_E/\lam_{E,2}$ emerging at the best-fit point are very different for the two cases.

We summarize the main characteristics of the scans of Model~1 and Model~2 in \reftable{tab:results_VL}.
In analogy to what we observed in the EFT analysis, the Bayes factor favors Model~1 over Model~2 by 5:1, which reads ``substantial'' evidence on Jeffrey's scale.

\begin{table*}[t]
	\begin{center}
		\begin{tabular*}{0.6\textwidth}{c|c|ccc|ccc}
        \toprule                     
        \bf $Z'$ + VL & $-$\textbf{ln} $\mathcal{Z}$ & \bf Pull & $\bm{\chi^2_{\textrm{TOT}}}$ & $\bm{\frac{\chi^2_{\textrm{TOT}}}{d.o.f}}$ &  \footnotesize  $\bm{m_{Z'}/g_X}$ & $\bm{M_{Q}/\lambda_Q}$ & $\bm{M_{\textrm{VL}}/\lambda_{\textrm{VL}}}$\\
		\toprule
            \bf Model 1 & $78.4$ & $4.5\,\sigma$ & $145.5$ & $1.05$ & $0.7\tev$ & $24.4\tev$ & $34.2\tev$\\
            \bf Model 2 & $80.0$ & $4.1\,\sigma$ & $149.4$ & $1.07$ & $0.7\tev$ &  $24.7\tev$ & $0.5\tev$ \\
            \toprule
		\end{tabular*}
		\caption{Evidence, pull to the SM, chi-squared statistics, and input parameters at the best-fit points of Model~1 and Model~2.}
		\label{tab:results_VL}
	\end{center}
\end{table*}

\subsubsection{A model with U(1)$_X$ charged VL leptons}

We finally consider an alternative to the $L_{\mu}-L_{\tau}$ model, obtained if one 
charges the VL leptons under the U(1)$_X$ symmetry, and leaves the SM leptons uncharged, see, e.g.,\cite{Sierra:2015fma}. 

\textbf{Model~3.} We add to the SM the following particle content
\bea
S:(\mathbf{1},\mathbf{1},0,-1)\,, & & \\
Q:(\mathbf{3},\mathbf{2},1/6,-1)\,, & \quad & Q':(\mathbf{\bar{3}},\mathbf{2},-1/6,1)\,,\\
L:(\mathbf{1},\mathbf{2},-1/2,1)\,, & \quad & L':(\mathbf{1},\mathbf{2},1/2,-1)\,.
\eea

The Lagrangian features terms involving the NP leptons,
\be
\mathcal{L}\supset -\lam_{L,i} S^{\ast} L' l_i-M_L L' L+\textrm{H.c.}\,,
\ee
where, again, $l_{i=1,2,3}$ are SM lepton doublets.

After rotating to the quark and lepton mass bases and again retaining only the Yukawa couplings relevant to $b\to s\mu\mu$ transitions one finds
\begin{multline}
C_9^{\mu}=-C_{10}^{\mu}=\frac{2\Lambda_v^2}{V_{tb}V_{ts}^{\ast}}\left(\frac{\lam_{Q,2}\lam_{Q,3}}{2 M_Q^2+\left(\lam_{Q,2}^2+\lam_{Q,3}^2\right)v_S^2}\right) \\
\times \left(\frac{\lam_{L,2}^2 v_S^2}{2 M_L^2+\lam_{L,2}^2 v_S^2}\right)\,.
\end{multline}
This model can be parametrized in terms of 3 parameters: $m_{Z'}/g_X$, 
$M_Q/\lam_Q$, and the hierarchical $\epsilon$, defined such that 
$M_L/\lam_{L,2}=\epsilon M_Q/\lam_Q$.

The $2\,\sigma$ regions of the 1-dimensional marginalized posterior pdf and of the profile likelihood in this case coincide and read 
\be
C_9^{\mu}=-C_{10}^{\mu}\in\left(-0.68,-0.29\right)\label{2sigma1dim}\,.
\ee
We apply this bound to Model~3, together with the bound from $B_s$ mixing. The favored $2\,\sigma$ regions are shown in 
\reffig{fig:mod}(c), with different color code for different $\epsilon$. The severe bound on $R_{BB}$ limits this model to strong hierarchies between VL quark and lepton masses. 

\subsection{Leptoquarks}

A second, well known class of models that can easily generate NP contributions to the Wilson coefficients of the EFT are leptoquarks. 
Much work has been done in the past few years 
on the phenomenology of leptoquarks in relation to the flavor anomalies, for some early references, 
see, e.g.,\cite{Gripaios:2014tna,Fajfer:2015ycq,Varzielas:2015iva,DiLuzio:2017vat,Hiller:2014yaa,Calibbi:2015kma,Bauer:2015knc,Barbieri:2015yvd}. 
Here, we limit ourselves to the analysis of but one of these models, the scalar SU(2) triplet $S_3$\cite{Hiller:2016kry,Dorsner:2016wpm,Dorsner:2017ufx,Crivellin:2017zlb,Hiller:2017bzc,Hiller:2018wbv,Becirevic:2018afm,Angelescu:2018tyl}, 
which can generate a $C_9^{\mu}=-C_{10}^{\mu}$ contribution at the tree level, like Model~3 of the previous subsection.

We thus introduce the scalar leptoquark
\be
S_3:(\mathbf{\bar{3}},\mathbf{3},1/3)\,.
\ee

The Lagrangian acquires a Yukawa term of the type
\be
\mathcal{L}\supset Y_{ij}q^T_i(i\sigma_2)S_3 l_j+\textrm{H.c.}\,,
\ee
where $q_i,\,l_j$ are the SM quark and lepton SU(2) doublets, 
$i,j=1,2,3$ span the SM generations and $S_3$ is a matrix
\be
S_3=\left( {\begin{array}{cc}
S_{1/3} & \sqrt{2}S_{4/3}\\
\sqrt{2}S_{-2/3} & -S_{1/3}
 \end{array} } \right),
\ee
with electric charges indicated in the subscripts.

After rotating to the mass basis one writes
\begin{multline}
\mathcal{L}\supset  -\hat{Y}_{ij} \left(S_{1/3}\,d_{L\,i} \nu_{L\,j}+\sqrt{2}S_{4/3}\,d_{L\,i} e_{L\,j}\right)\\
 -\widetilde{Y}_{ij} 
\left(S_{1/3}\,u_{L\,i} e_{L\,j}-\sqrt{2}S_{-2/3}\,u_{L\,i} \nu_{L\,j}\right)+\textrm{H.c.}\,,
\end{multline}
where $\widetilde{Y}_{ij}^T=\hat{Y}_{ik}^T(V_{\textrm{CKM}}^{\dag})_{kj}$. 
Note that couplings of the type $qS_3^{\dag}q$, which are very dangerous for proton decay, 
are allowed in the SM, so that in UV complete models one should make sure they are forbidden
by an additional symmetry.

By matching to the EFT one finds 
\be
C_9^{\mu}=-C_{10}^{\mu}=\frac{\pi v_h^2}{V_{tb}V_{ts}^{\ast}\,\alpha_{\textrm{em}}}\frac{\hat{Y}_{b\mu}\hat{Y}_{s\mu}^{\ast}}{m_{S_3}^2}\,,
\ee
where in this and what follows we have assumed that the mass of the triplet states is the same, $m_{S_3}$.

The constraint from the 1-dimensional EFT at $2\,\sigma$ is given in \refeq{2sigma1dim}.
This leads to
\be\label{lepto}
0.4\times 10^{-3} \left(\frac{m_{S_3}}{\tev}\right)^2 \leq\hat{Y}_{b\mu}\hat{Y}_{s\mu}^{\ast}\leq 1.1\times 10^{-3} \left(\frac{m_{S_3}}{\tev}\right)^2\,.
\ee
 
If one starts with $\hat{Y}_{b\mu}\hat{Y}_{s\mu}^{\ast}\neq 0$, the 
CKM matrix generates additional Yukawa couplings,
\bea
\widetilde{Y}_{u\mu}&=&(V_{\textrm{CKM}}^{\ast})_{12}\hat{Y}_{22}+(V_{\textrm{CKM}}^{\ast})_{13}\hat{Y}_{32}\,,\nonumber\\
\widetilde{Y}_{c\mu}&=&(V_{\textrm{CKM}}^{\ast})_{22}\hat{Y}_{22}+(V_{\textrm{CKM}}^{\ast})_{23}\hat{Y}_{32}\,,\nonumber\\
\widetilde{Y}_{t\mu}&=&(V_{\textrm{CKM}}^{\ast})_{32}\hat{Y}_{22}+(V_{\textrm{CKM}}^{\ast})_{33}\hat{Y}_{32}\,,
\eea
so that possible complementary constraints come from $B\to K^{(\ast)}\nu\bar{\nu}$, $b\to c \mu^-\bar{\nu}$ decays, 
$t\to c\,\mu^+\mu^-$, and $t\to c\,\nu\nu$.

The most dangerous constraint is possibly given by $B\to K^{(\ast)}\nu\bar{\nu}$ decay.
The bound can be expressed as\cite{Buras:2014fpa,Fajfer:2015ycq}
\begin{multline}
\textrm{Br}_{(\textrm{SM})}\cdot\left[1+\frac{4 \pi v_h^2}{3\alpha V_{tb}V_{ts}^{\ast}m_{S_3}^2 C_L^{\textrm{SM}}}\Re(\hat{Y}_{s\mu}\hat{Y}_{b\mu}^{\ast})\right.\\
\left.+\,\frac{1}{3|C_L^{\textrm{SM}}|^2}
\left(\frac{2\pi v^2}{\alpha V_{tb}V_{ts}^{\ast}m_{S_3}^2}\right)^2 |\hat{Y}_{s\mu}|^2 |\hat{Y}_{b\mu}|^2\right]\leq \textrm{Br}_{(\textrm{90\%~CL})}\,,
\end{multline}
where $\textrm{Br}_{(\textrm{SM})}=(4.0\pm 0.5)\times 10^{-6}$, $\textrm{Br}_{(\textrm{90\%~CL})}=1.6 \times 10^{-5}$, and $C_L^{\textrm{SM}}=-6.38\pm 0.06$.
We get the limit
\be
\Re(\hat{Y}_{b\mu}\hat{Y}_{s\mu}^{\ast})\lesssim 2.2\times 10^{-2}\left(\frac{m_{S_3}}{\tev}\right)^2\,,
\ee
which does not constrain the parameter space emerging in \refeq{lepto}.

\section{Summary and conclusions\label{sec:summary}}

In this paper we have presented a global Bayesian analysis of the NP effects on effective operators
of semileptonic $b\to s$ transitions after the very recent updated measurement of $R_K$ at LHCb and new results for the observable $R_{K^*}$ in $B^0$-meson decays, as well as the first measurement of its counterpart  
$R_{K^{*+}}$ in $B^+$ decays at Belle. We have performed global fits with 1, 2, 4, and 8 Wilson coefficients as inputs, 
plus one CKM nuisance parameter to take into account 
uncertainties that are not factorizable with the NP effects. From the fits, we then inferred 
the 68\% and 95.4\% credibility regions of the marginalized posterior probability density for all models.

The new measurement of $R_K$ is closer in central value to the SM prediction than the Run~1 determination, but 
the much improved precision of the new data keeps it at $2.5\,\sigma$ from the SM. 
As a result the high-probability region of the posterior pdf in the NP Wilson coefficients $C_9^{\mu}$ and $C_{10}^{\mu}$ shifts slightly towards the zero value with respect to the 
scans with the Run~1 determination of $R_K$, but the overall pull remains quite large, at the level of 4--5~$\sigma$, 
quite independently of the number of scanned input coefficients.

We have confirmed previous observations that the impact of the Wilson coefficients of the electron sector on the data is negligible with respect to the muon sector. Moreover, 
a pair-like comparison of the Bayes factors of different models has allowed us to determine that the two scans characterized by the inputs 
$C_9^{\mu}$, $C_{9}^{\prime \mu}$, and $C_9^{\mu}$, $C_{10}^{\mu}$,
$C_{9}^{\prime \mu}$, $C_{10}^{\prime \mu}$ are favored by the data, with respect to all other combinations.
The frequentist measures of the goodness of fit like the 
minimum chi-squared per degree of freedom confirm this preference.
 
Finally, we have also analyzed a few well-known BSM models that can provide a high energy framework for the EFT analysis.
These include the exchange of a heavy $Z^{'}$ gauge boson in models with heavy vector-like fermions and a scalar field whose vev breaks spontaneously the new symmetry,
and a model with scalar leptoquarks. Despite the introduction of new constraints that are specific to the model-dependent analysis, when 
it comes to determining which hypotheses are strongly favored by the data, the Bayes factors mirror 
the results of the EFT fits, i.e., models that can generate the $C_9^{\mu}$ and $C_{9}^{\prime \mu}$ Wilson coefficients after integrating out heavy degrees of freedom are 
preferred with respect to other combinations.

\begin{acknowledgements}
We would like to thank Andrew Fowlie for useful correspondence on \texttt{Superplot}. 
KK and DK are supported in part by the National Science Centre (Poland) under the research Grant No. 2017/26/E/ST2/00470.
EMS is supported in part by the National Science Centre (Poland) under the research Grant No. 2017/26/D/ST2/00490. 
The use of the CIS computer cluster at the National Centre for Nuclear Research in Warsaw is gratefully acknowledged.
\end{acknowledgements}

\appendix

\section{List of observables used in the global analysis}\label{app:obs}

In this appendix we provide a tabularized list of all the observables included in our global analysis as components of the likelihood function (Tables~\ref{tab:data_B0_exp0}-\ref{tab:data_B0_exp18}). 
For each of them we show the experimental measurement and the SM prediction derived with \flavio, 
which includes the theoretical error obtained by calculating the spread of values for a given observable, when a set of input parameters (form factors, bag parameters, decay constants, masses of the particles) 
were randomly generated for 2000 times. In the last column we also present a deviation of the measurement from the SM prediction that quantifies the significance of a potential anomaly.

\begin{table}[h]
	\begin{center}\footnotesize
		\begin{tabular}{l|c|c|c}
			\hline\hline 
			\multicolumn{4}{c}{LFUV observables} \\
		    \hline\hline
            Observable & SM prediction & Experimental value & Deviation\\
			\hline
			\multicolumn{4}{c}{LHCb ($B^+\to K^+l^+l^-$)\cite{Aaij:2019wad}}\\
            \hline
            $R_{K}^{[1.1,6]}$ & $1.001\pm 0.000$ & $0.846^{+0.060}_{-0.054}\pm 0.016$ & $\bm{2.5\sigma}$ \\
            \hline
			\multicolumn{4}{c}{LHCb ($B^0\to K^{*0}l^+l^-$)\cite{Aaij:2017vbb}}\\
            \hline
            $R_{K*}^{[0.045,1.1]}$ & $0.928\pm 0.004$ & $0.660^{+0.110}_{-0.070}\pm 0.024$ & $\bm{2.4\sigma}$ \\
            $R_{K*}^{[1.1,6]}$ & $0.997\pm 0.001$ & $0.685^{+0.113}_{-0.069}\pm 0.047$ & $\bm{2.5\sigma}$ \\
            \hline
			\multicolumn{4}{c}{Belle ($B^0\to K^{*0}l^+l^-$)\cite{Abdesselam:2019wac}}\\
            \hline
            $R_{K*}^{[0.045,1.1]}$ & $0.928\pm 0.004$ & $0.46^{+0.55}_{-0.27}\pm 0.07$ & $0.8\sigma$ \\
            $R_{K*}^{[1.1,6]}$ & $0.997\pm 0.001$ & $1.06^{+0.63}_{-0.38}\pm 0.13$ & $0.2\sigma$ \\            
            $R_{K*}^{[15,19]}$ & $0.997\pm 0.000$ & $1.12^{+0.61}_{-0.36}\pm 0.1$ & $0.3\sigma$ \\ 
            \hline
			\multicolumn{4}{c}{Belle ($B^+\to K^{*+}l^+l^-$)\cite{Abdesselam:2019wac}}\\
            \hline
            $R_{K*}^{[0.045,1.1]}$ & $0.928\pm 0.004$ & $0.62^{+0.60}_{-0.36}\pm 0.10$ & $0.5\sigma$ \\
            $R_{K*}^{[1.1,6]}$ & $0.997\pm 0.001$ & $0.72^{+0.99}_{-0.44}\pm 0.18$ & $0.3\sigma$ \\            
            $R_{K*}^{[15,19]}$ & $0.998\pm 0.000$ & $1.40^{+1.99}_{-0.68}\pm 0.11$ & $0.6\sigma$ \\ 
            \hline
		\end{tabular}
		\caption{\footnotesize LFUV observables included in the global fit.}
		\label{tab:data_B0_exp0} 
	\end{center}
\end{table}

\begin{table}[p]
	\begin{center}\footnotesize
		\begin{tabular}{l|c|c|c}
			\hline\hline
			\multicolumn{4}{c}{$B^0\to K^{*0}\mu^+\mu^-$ angular observables} \\
			\hline\hline
            Observable & SM prediction & Experimental value & Pull\\
			\hline
			\multicolumn{4}{c}{LHCb\cite{Aaij:2015oid}}\\
            \hline
            $\langle F_L\rangle^{[1.1,2.5]}$ & $0.761\pm 0.044$ & $0.666^{+0.083}_{-0.077}\pm 0.022$ & $1.0\sigma$ \\
            $\langle F_L\rangle^{[2.5,4]}$ & $0.796\pm 0.036$ & $0.876^{+0.109}_{-0.097}\pm 0.017$ & $0.7\sigma$ \\
            $\langle F_L\rangle^{[4,6]}$ & $0.711\pm 0.049$ & $0.611^{+0.052}_{-0.053}\pm 0.017$ & $1.4\sigma$ \\
            $\langle F_L\rangle^{[15,19]}$ & $0.340\pm 0.022$ & $0.344^{+0.028}_{-0.030}\pm 0.008$ & $0.1\sigma$ \\    
            $\langle A_{FB}\rangle^{[1.1,2.5]}$ & $-0.137\pm 0.030$ & $-0.191^{+0.068}_{-0.080}\pm 0.012$ & $0.6\sigma$ \\
            $\langle A_{FB}\rangle^{[2.5,4]}$ & $-0.017\pm 0.032$ & $-0.118^{+0.082}_{-0.090}\pm 0.007$ & $1.1\sigma$ \\
            $\langle A_{FB}\rangle^{[4,6]}$ & $0.123\pm 0.042$ & $0.025^{+0.051}_{-0.052}\pm 0.004$ & $1.5\sigma$ \\
            $\langle A_{FB}\rangle^{[15,19]}$ & $0.368\pm 0.021$ & $0.355^{+0.027}_{-0.027}\pm 0.009$ & $0.4\sigma$ \\            
            $\langle S_3\rangle^{[1.1,2.5]}$ & $0.002\pm 0.005$ & $-0.077^{+0.087}_{-0.105}\pm 0.005$ & $0.8\sigma$ \\
            $\langle S_3\rangle^{[2.5,4]}$ & $-0.011\pm 0.004$ & $0.035^{+0.098}_{-0.089}\pm 0.007$ & $0.5\sigma$ \\
            $\langle S_3\rangle^{[4,6]}$ & $-0.025\pm 0.009$ & $0.035^{+0.069}_{-0.068}\pm 0.007$ & $0.9\sigma$ \\
            $\langle S_3\rangle^{[15,19]}$ & $-0.205\pm 0.016$ & $-0.163^{+0.033}_{-0.033}\pm 0.009$ & $1.1\sigma$ \\ 
            $\langle S_4\rangle^{[1.1,2.5]}$ & $ -0.026\pm 0.017$ & $-0.077^{+0.111}_{-0.113}\pm 0.005$ & $0.4\sigma$ \\
            $\langle S_4\rangle^{[2.5,4]}$ & $-0.152\pm 0.022$ & $-0.234^{+0.127}_{-0.144}\pm 0.006$ & $0.6\sigma$ \\
            $\langle S_4\rangle^{[4,6]}$ & $-0.224\pm 0.020$ & $-0.219^{+0.086}_{-0.084}\pm 0.008$ & $0.1\sigma$ \\
            $\langle S_4\rangle^{[15,19]}$ & $-0.300\pm 0.006 $ & $-0.284^{+0.038}_{-0.041}\pm 0.007$ & $0.4\sigma$ \\ 
            $\langle S_5\rangle^{[1.1,2.5]}$ & $0.053\pm 0.035$ & $0.137^{+0.099}_{-0.094}\pm 0.009$ & $0.8\sigma$ \\
            $\langle S_5\rangle^{[2.5,4]}$ & $-0.194\pm 0.039$ & $-0.022^{+0.110}_{-0.103}\pm 0.008$ & $1.5\sigma$ \\
            $\langle S_5\rangle^{[4,6]}$ & $-0.337\pm 0.035$ & $-0.146^{+0.077}_{-0.078}\pm 0.011$ & $\bm{2.2\sigma}$ \\
            $\langle S_5\rangle^{[15,19]}$ & $-0.281\pm 0.017$ & $-0.325^{+0.036}_{-0.037}\pm 0.009$ & $1.1\sigma$ \\ 
            $\langle S_7\rangle^{[1.1,2.5]}$ & $-0.027\pm 0.030$ & $-0.219^{+0.094}_{-0.104}\pm 0.004$ & $1.8\sigma$ \\
            $\langle S_7\rangle^{[2.5,4]}$ & $-0.020\pm 0.041$ & $0.068^{+0.120}_{-0.112}\pm 0.005$ & $0.7\sigma$ \\
            $\langle S_7\rangle^{[4,6]}$ & $-0.013\pm 0.051$ & $-0.016^{+0.081}_{-0.080}\pm 0.004$ & $0.0\sigma$ \\
            $\langle S_7\rangle^{[15,19]}$ & $-0.001\pm 0.001$ & $0.048^{+0.043}_{-0.043}\pm 0.006$ & $1.1\sigma$ \\
            $\langle S_8\rangle^{[1.1,2.5]}$ & $-0.007\pm 0.013$ & $-0.098^{+0.108}_{-0.123}\pm 0.005$ & $0.7\sigma$ \\
            $\langle S_8\rangle^{[2.5,4]}$ & $-0.006\pm 0.014$ & $0.030^{+0.129}_{-0.131}\pm 0.006$ & $0.3\sigma$ \\
            $\langle S_8\rangle^{[4,6]}$ & $-0.005\pm 0.015$ & $0.167^{+0.094}_{-0.091}\pm 0.004$ & $1.8\sigma$ \\
            $\langle S_8\rangle^{[15,19]}$ & $0.000\pm 0.000$ & $0.028^{+0.044}_{-0.045}\pm 0.003$ & $0.6\sigma$ \\ 
            $\langle S_9\rangle^{[1.1,2.5]}$ & $-0.001\pm 0.005$ & $-0.119^{+0.087}_{-0.104}\pm 0.005$ & $1.1\sigma$ \\
            $\langle S_9\rangle^{[2.5,4]}$ & $-0.001\pm 0.002$ & $-0.092^{+0.105}_{-0.125}\pm 0.007$ & $0.7\sigma$ \\
            $\langle S_9\rangle^{[4,6]}$ & $-0.001\pm 0.005$ & $-0.032^{+0.071}_{-0.071}\pm 0.004$ & $0.4\sigma$ \\
            $\langle S_9\rangle^{[15,19]}$ & $0.000\pm 0.000$ & $-0.053^{+0.039}_{-0.039}\pm 0.002$ & $1.4\sigma$ \\  
            \hline
			\multicolumn{4}{c}{Belle\cite{Wehle:2016yoi}}\\
		    \hline
		    $\langle P_4^{'}\rangle^{[0.1,4]}$ & $-0.03\pm 0.03$ & $-0.38^{+0.50}_{-0.48}\pm 0.12$ & $0.7\sigma$ \\
            $\langle P_4^{'}\rangle^{[14.18,19]}$ & $-0.63\pm 0.01$ & $-0.10^{+0.39}_{-0.39}\pm 0.07$ & $1.3\sigma$ \\ 
 		    $\langle P_5^{'}\rangle^{[0.1,4]}$ & $0.15\pm 0.06$ & $0.42^{+0.39}_{-0.39}\pm 0.14$ & $0.6\sigma$ \\
            $\langle P_5^{'}\rangle^{[14.18,19]}$ & $-0.63\pm 0.03$ & $-0.13^{+0.39}_{-0.35}\pm 0.06$ & $1.3\sigma$ \\
            \hline
			\end{tabular}
		\caption{Angular observables of $B^0\to K^{*0}\mu^+\mu^-$ included in the global fit.}
		\label{tab:data_B0_exp1} 
	\end{center}
\end{table}

\newpage
\begin{table}[p]
	\begin{center}\footnotesize
		\begin{tabular}{l|c|c|c}
			\hline\hline 
			\multicolumn{4}{c}{$B^0\to K^{*0}\mu^+\mu^-$ angular observables} \\
		    \hline\hline
            Observable & SM prediction & Experimental value & Pull\\
			\hline
			\multicolumn{4}{c}{ATLAS\cite{Aaboud:2018krd}}\\
            \hline
            $\langle F_L\rangle^{[0.04,2]}$ & $0.39\pm 0.06$ & $0.44\pm 0.08\pm 0.07$ & $0.4\sigma$ \\
            $\langle F_L\rangle^{[2,4]}$ & $0.80\pm 0.04$ & $0.64\pm 0.11\pm 0.05$ & $1.3\sigma$ \\
            $\langle F_L\rangle^{[4,6]}$ & $0.71\pm 0.05$ & $0.42\pm 0.13\pm 0.12$ & $1.6\sigma$ \\
            $\langle S_3\rangle^{[0.04,2]}$ & $0.01\pm 0.01$ & $-0.02\pm 0.09\pm 0.02$ & $0.3\sigma$ \\
            $\langle S_3\rangle^{[2,4]}$ & $-0.01\pm 0.00$ & $-0.15\pm 0.10\pm 0.07$ & $1.2\sigma$ \\
            $\langle S_3\rangle^{[4,6]}$ & $-0.02\pm 0.01$ & $0.00\pm 0.12\pm 0.07$ & $0.2\sigma$ \\   
            $\langle S_4\rangle^{[0.04,2]}$ & $0.06\pm 0.01$ & $0.15\pm 0.20\pm 0.10$ & $0.4\sigma$ \\
            $\langle S_4\rangle^{[2,4]}$ & $-0.13\pm 0.02$ & $-0.37\pm 0.15\pm 0.10$ & $1.3\sigma$ \\
            $\langle S_4\rangle^{[4,6]}$ & $-0.22\pm 0.02$ & $0.32\pm 0.16\pm 0.09$ & $\bm{3.0\sigma}$ \\    
            $\langle S_5\rangle^{[0.04,2]}$ & $0.20\pm 0.01$ & $0.33\pm 0.13\pm 0.08$ & $0.9\sigma$ \\
            $\langle S_5\rangle^{[2,4]}$ & $-0.16\pm 0.04$ & $-0.16\pm 0.15\pm 0.06$ & $0.0\sigma$ \\
            $\langle S_5\rangle^{[4,6]}$ & $-0.34\pm 0.04$ & $0.13\pm 0.18\pm 0.09$ & $\bm{2.3\sigma}$ \\               
            $\langle S_7\rangle^{[0.04,2]}$ & $-0.02\pm 0.02$ & $-0.09\pm 0.10\pm 0.02$ & $0.7\sigma$ \\
            $\langle S_7\rangle^{[2,4]}$ & $-0.02\pm 0.04$ & $0.15\pm 0.14\pm 0.09$ & $1.0\sigma$ \\
            $\langle S_7\rangle^{[4,6]}$ & $-0.01\pm 0.05$ & $0.03\pm 0.13\pm 0.07$ & $0.3\sigma$ \\  
            $\langle S_8\rangle^{[0.04,2]}$ & $0.00\pm 0.01$ & $-0.14\pm 0.24\pm 0.09$ & $0.5\sigma$ \\
            $\langle S_8\rangle^{[2,4]}$ & $-0.01\pm 0.01$ & $0.52\pm 0.20\pm 0.19$ & $1.9\sigma$ \\
            $\langle S_8\rangle^{[4,6]}$ & $0.00\pm 0.02$ & $-0.12\pm 0.21\pm 0.05$ & $0.5\sigma$ \\            
			\hline
			\multicolumn{4}{c}{CMS 2015\cite{Khachatryan:2015isa}}\\
            \hline
            $\langle F_L\rangle^{[1,2]}$ & $0.73\pm 0.05$ & $0.64^{+0.10}_{-0.09}\pm 0.07$ & $0.7\sigma$ \\
            $\langle F_L\rangle^{[2,4.3]}$ & $0.79\pm 0.04$ & $0.80^{+0.08}_{-0.08}\pm 0.06$ & $0.1\sigma$ \\
            $\langle F_L\rangle^{[4.3,6]}$ & $0.70\pm 0.05$ & $0.62^{+0.10}_{-0.09}\pm 0.07$ & $0.6\sigma$ \\
            $\langle A_{FB}\rangle^{[1,2]}$ & $-0.16\pm 0.03$ & $-0.27^{+0.17}_{-0.40}\pm 0.07$ & $0.3\sigma$ \\
            $\langle A_{FB}\rangle^{[2,4.3]}$ & $-0.02\pm 0.03$ & $-0.12^{+0.15}_{-0.17}\pm 0.05$ & $0.5\sigma$ \\
            $\langle A_{FB}\rangle^{[4.3,6]}$ & $0.13\pm 0.04$ & $0.01^{+0.15}_{-0.15}\pm 0.03$ & $0.8\sigma$ \\
            \hline
			\multicolumn{4}{c}{CMS 2017\cite{Sirunyan:2017dhj}}\\
            \hline
            $\langle P_1\rangle^{[1,2]}$ & $0.05\pm 0.05$ & $0.12^{+0.46}_{-0.47}\pm 0.10$ & $0.2\sigma$ \\
            $\langle P_1\rangle^{[2,4.3]}$ & $-0.11\pm 0.04$ & $-0.69^{+0.58}_{-0.27}\pm 0.23$ & $1.0\sigma$ \\
            $\langle P_1\rangle^{[4.3,6]}$ & $-0.18\pm 0.05$ & $0.53^{+0.24}_{-0.33}\pm 0.19$ & $1.9\sigma$ \\
            $\langle P_5^{'}\rangle^{[1,2]}$ & $0.29\pm 0.07$ & $0.10^{+0.32}_{-0.31}\pm 0.07$ & $0.5\sigma$ \\
            $\langle P_5^{'}\rangle^{[2,4.3]}$ & $-0.45\pm 0.10$ & $-0.57^{+0.34}_{-0.31}\pm 0.18$ & $0.3\sigma$ \\
            $\langle P_5^{'}\rangle^{[4.3,6]}$ & $-0.77\pm 0.08$ & $-0.96^{+0.22}_{-0.21}\pm 0.25$ & $0.7\sigma$ \\
            \hline
		\end{tabular}
		\caption{Angular observables of $B^0\to K^{*0}\mu^+\mu^-$ included in the global fit.}
		\label{tab:data_B0_exp2} 
	\end{center}
\end{table}

\begin{table}[p]
	\begin{center}\footnotesize
		\begin{tabular}{l|c|c|c}
			\hline\hline
			\multicolumn{4}{c}{$B^0\to K^{*0}\mu^+\mu^-$ angular observables} \\
		    \hline\hline
            Observable & SM prediction & Experimental value & Pull\\
            \hline
			\multicolumn{4}{c}{CDF\cite{CDF}}\\
		    \hline
            $\langle F_L\rangle^{[0,2]}$ & $0.39\pm 0.06$ & $0.26^{+0.14}_{-0.13}\pm 0.04$ & $0.8\sigma$ \\
            $\langle F_L\rangle^{[2,4.3]}$ & $0.79\pm 0.04$ & $0.72^{+0.15}_{-0.17}\pm 0.09$ & $0.4\sigma$ \\
            $\langle A_{FB}\rangle^{[0,2]}$ & $-0.10\pm 0.01$ & $0.07^{+0.29}_{-0.28}\pm 0.11$ & $0.6\sigma$ \\
            $\langle A_{FB}\rangle^{[2,4.3]}$ & $-0.03\pm 0.03$ & $-0.11^{+0.34}_{-0.45}\pm 0.16$ & $0.2\sigma$ \\
			\hline
		\end{tabular}
		\caption{Angular observables of $B^0\to K^{*0}\mu^+\mu^-$ included in the global fit.}
		\label{tab:data_B0_exp3} 
	\end{center}
\end{table}

\begin{table}[p]
	\begin{center}\footnotesize
		\begin{tabular}{l|c|c|c}
			\hline\hline
			\multicolumn{4}{c}{$B^0\to K^{*0}\mu^+\mu^-$ differential branching ratio} \\
		    \hline\hline
            Observable & SM prediction & Experimental value & Pull\\
            \hline
			\multicolumn{4}{c}{CDF\cite{CDF}}\\
		    \hline
            $10^8\times\langle \frac{d\textrm{BR}}{dq^2}\rangle^{[0,2]}$ & $8.23\pm 1.14$ & $9.12\pm 1.73\pm 0.49$ & $0.4\sigma$ \\
            $10^8\times\langle \frac{d\textrm{BR}}{dq^2}\rangle^{[2,4.3]}$ & $4.50\pm 0.69$ & $4.61\pm 1.19\pm 0.27$ & $0.1\sigma$ \\
            \hline
			\multicolumn{4}{c}{CMS\cite{Khachatryan:2015isa}}\\
		    \hline
            $10^8\times\langle \frac{d\textrm{BR}}{dq^2}\rangle^{[1,2]}$ & $4.9\pm 0.7$ & $4.6\pm 0.7\pm 0.3$ & $0.2\sigma$ \\
            $10^8\times\langle \frac{d\textrm{BR}}{dq^2}\rangle^{[2,4.3]}$ & $4.5\pm 0.7$ & $3.3\pm 0.5\pm 0.2$ & $1.4\sigma$ \\
            $10^8\times\langle \frac{d\textrm{BR}}{dq^2}\rangle^{[4.3,6]}$ & $5.1\pm 0.8$ & $3.4\pm 0.5\pm 0.3$ & $1.7\sigma$ \\
            \hline
			\multicolumn{4}{c}{LHCb\cite{Aaij:2016flj}}\\
		    \hline
            $10^8\times \langle \frac{d\textrm{BR}}{dq^2}\rangle^{[1.1,2.5]}$ & $4.65\pm 0.68$ & $3.26^{+0.32}_{-0.31}\pm 0.10\pm0.22$ & $1.8\sigma$ \\
            $10^8\times\langle \frac{d\textrm{BR}}{dq^2}\rangle^{[2.5,4]}$ & $4.49\pm 0.69$ & $3.34^{+0.31}_{-0.33}\pm 0.09\pm0.23$ & $1.4\sigma$ \\
            $10^8\times\langle \frac{d\textrm{BR}}{dq^2}\rangle^{[4,6]}$ & $5.02\pm 0.76$ & $3.54^{+0.27}_{-0.26}\pm 0.09\pm 0.24$ & $1.8\sigma$ \\
            $10^8\times\langle \frac{d\textrm{BR}}{dq^2}\rangle^{[15,19]}$ & $5.95\pm 0.63$ & $4.36^{+0.18}_{-0.19}\pm 0.07\pm 0.30$ & $\bm{2.2\sigma}$ \\
			\hline
		\end{tabular}
		\caption{Binned differential branching ratio of $B^0\to K^{*0}\mu^+\mu^-$ included in the global fit.}
		\label{tab:data_B0_exp4} 
	\end{center}
\end{table}

\begin{table}[p]
	\begin{center}\footnotesize
		\begin{tabular}{l|c|c|c}
			\hline\hline
			\multicolumn{4}{c}{$B^0\to K^{0}\mu^+\mu^-$ differential branching ratio} \\
		    \hline\hline
            Observable & SM prediction & Experimental value & Pull\\
            \hline
			\multicolumn{4}{c}{CDF\cite{CDF}}\\
		    \hline
            $10^8\times\langle \frac{d\textrm{BR}}{dq^2}\rangle^{[0,2]}$ & $3.28\pm 0.57$ & $2.45\pm 1.59\pm 0.21$ & $0.5\sigma$ \\
            $10^8\times\langle \frac{d\textrm{BR}}{dq^2}\rangle^{[2,4.3]}$ & $3.25\pm 0.56$ & $2.55\pm 1.70\pm 0.35$ & $0.4\sigma$ \\
            \hline
			\multicolumn{4}{c}{LHCb\cite{Aaij:2014pli}}\\
		    \hline
            $10^8\times\langle \frac{d\textrm{BR}}{dq^2}\rangle^{[0.1,2]}$ & $3.28\pm 0.57$ & $1.22^{+0.59}_{-0.52}\pm 0.06$ & $\bm{2.5\sigma}$ \\
            $10^8\times\langle \frac{d\textrm{BR}}{dq^2}\rangle^{[2,4]}$ & $3.25\pm 0.55$ & $1.87^{+0.55}_{-0.49}\pm 0.09$ & $1.7\sigma$ \\
            $10^8\times\langle \frac{d\textrm{BR}}{dq^2}\rangle^{4,6]}$ & $3.21\pm 0.54$ & $1.73^{+0.53}_{-0.48}\pm 0.09$ & $1.9\sigma$ \\
            $10^8\times\langle \frac{d\textrm{BR}}{dq^2}\rangle^{15,22]}$ & $1.39\pm 0.16$ & $0.95^{+0.16}_{-0.15}\pm 0.05$ & $1.9\sigma$ \\
            \hline
		\end{tabular}
		\caption{Binned differential branching ratio of $B^0\to K^{0}\mu^+\mu^-$ included in the global fit.}
		\label{tab:data_B0_exp5} 
	\end{center}
\end{table}

\begin{table}[p]
	\begin{center}\footnotesize
		\begin{tabular}{l|c|c|c}
			\hline\hline
			\multicolumn{4}{c}{$B^+\to K^{+}\mu^+\mu^-$ differential branching ratio} \\
		    \hline\hline
            Observable & SM prediction & Experimental value & Pull\\
            \hline
			\multicolumn{4}{c}{CDF\cite{CDF}}\\
		    \hline
            $10^8\times\langle \frac{d\textrm{BR}}{dq^2}\rangle^{[0,2]}$ & $3.52\pm 0.61$ & $1.80\pm 0.53\pm 0.12$ & $\bm{2.1\sigma}$ \\
            $10^8\times\langle \frac{d\textrm{BR}}{dq^2}\rangle^{[2,4.3]}$ & $3.50\pm 0.59$ & $3.16\pm 0.54\pm 0.18$ & $0.4\sigma$ \\
            \hline
			\multicolumn{4}{c}{LHCb\cite{Aaij:2014pli}}\\
		    \hline
            $10^8\times\langle \frac{d\textrm{BR}}{dq^2}\rangle^{[1.1,2]}$ & $3.53\pm 0.61$ & $2.33\pm 0.15\pm 0.12$ & $1.9\sigma$ \\
            $10^8\times\langle \frac{d\textrm{BR}}{dq^2}\rangle^{[2,3]}$ & $3.51\pm 0.61$ & $2.82\pm 0.16\pm 0.14$ & $1.1\sigma$ \\
            $10^8\times\langle \frac{d\textrm{BR}}{dq^2}\rangle^{[3,4]}$ & $3.49\pm 0.59$ & $2.54\pm 0.15\pm 0.13$ & $1.5\sigma$ \\
            $10^8\times\langle \frac{d\textrm{BR}}{dq^2}\rangle^{[4,5]}$ & $3.47\pm 0.59$ & $2.21\pm 0.14\pm 0.11$ & $\bm{2.0\sigma}$ \\
            $10^8\times\langle \frac{d\textrm{BR}}{dq^2}\rangle^{[5,6]}$ & $3.45\pm 0.57$ & $2.31\pm 0.14\pm 0.12$ & $1.9\sigma$ \\
            $10^8\times\langle \frac{d\textrm{BR}}{dq^2}\rangle^{[15,22]}$ & $1.51\pm 0.17$ & $1.21\pm 0.04\pm 0.06$ & $1.6\sigma$ \\
            \hline
		\end{tabular}
		\caption{Binned differential branching ratio of $B^+\to K^{+}\mu^+\mu^-$ included in the global fit.}
		\label{tab:data_B0_exp6} 
	\end{center}
\end{table}

\begin{table}[p]
	\begin{center}\footnotesize
		\begin{tabular}{l|c|c|c}
			\hline\hline
			\multicolumn{4}{c}{$B^+\to K^{*+}\mu^+\mu^-$ differential branching ratio} \\
		    \hline\hline
            Observable & SM prediction & Experimental value & Pull\\
            \hline
			\multicolumn{4}{c}{CDF\cite{CDF}}\\
		    \hline
            $10^8\times\langle \frac{d\textrm{BR}}{dq^2}\rangle^{[0,2]}$ & $8.63\pm 1.23$ & $7.50\pm 4.68\pm 0.88$ & $0.2\sigma$ \\
            $10^8\times\langle \frac{d\textrm{BR}}{dq^2}\rangle^{[2,4.3]}$ & $4.90\pm 0.74$ & $4.94\pm 3.58\pm 0.63$ & $0.0\sigma$ \\
            \hline
			\multicolumn{4}{c}{LHCb\cite{Aaij:2014pli}}\\
		    \hline
            $10^8\times\langle \frac{d\textrm{BR}}{dq^2}\rangle^{[0.1,2]}$ & $7.93\pm 1.09$ & $5.92^{+1.44}_{-1.30}\pm 0.40$ & $1.1\sigma$ \\
            $10^8\times\langle \frac{d\textrm{BR}}{dq^2}\rangle^{[2,4]}$ & $4.87\pm 0.73$ & $5.59^{+1.59}_{-1.44}\pm 0.38$ & $0.4\sigma$ \\
            $10^8\times\langle \frac{d\textrm{BR}}{dq^2}\rangle^{4,6]}$ & $5.43\pm 0.82$ & $2.49^{+1.10}_{-0.96}\pm 0.17$ & $\bm{2.1\sigma}$ \\
            $10^8\times\langle \frac{d\textrm{BR}}{dq^2}\rangle^{15,19]}$ & $6.42\pm 0.67$ & $3.95^{+0.80}_{-0.73}\pm 0.28$ & $\bm{2.3\sigma}$ \\
            \hline
		\end{tabular}
		\caption{Binned differential branching ratio of $B^+\to K^{*+}\mu^+\mu^-$ included in the global fit.}
		\label{tab:data_B0_exp7} 
	\end{center}
\end{table}

\begin{table}[p]
	\begin{center}\footnotesize
		\begin{tabular}{l|c|c|c}
			\hline\hline
			\multicolumn{4}{c}{$B_s^0\to \phi\mu^+\mu^-$ differential branching ratio} \\
		    \hline\hline
            Observable & SM prediction & Experimental value & Pull\\
            \hline
			\multicolumn{4}{c}{LHCb\cite{Aaij:2015esa}}\\
		    \hline
            $10^8\times\langle \frac{d\textrm{BR}}{dq^2}\rangle^{[1,6]}$ & $5.39\pm 0.65$ & $2.58^{+0.33}_{-0.31}\pm 0.08\pm 0.19$ & $\bm{3.7\sigma}$ \\
            $10^8\times\langle \frac{d\textrm{BR}}{dq^2}\rangle^{[15,19]}$ & $5.57\pm 0.47$ & $4.04^{+0.39}_{-0.38}\pm 0.13\pm 0.30$ & $\bm{2.2\sigma}$ \\
            \hline
		\end{tabular}
		\caption{Binned differential branching ratio of $B_s^0\to \phi\mu^+\mu^-$ included in the global fit.}
		\label{tab:data_B0_exp8} 
	\end{center}
\end{table}

\begin{table}[p]
	\begin{center}\footnotesize
		\begin{tabular}{l|c|c|c}
			\hline\hline
			\multicolumn{4}{c}{$B_s^0\to \phi\mu^+\mu^-$ angular observables} \\
		    \hline\hline
            Observable & SM prediction & Experimental value & Pull\\
            \hline
			\multicolumn{4}{c}{LHCb\cite{Aaij:2015esa}}\\
		    \hline
            $\langle F_L\rangle^{[0.1,2]}$ & $0.50\pm 0.04$ & $0.20^{+0.08}_{-0.09}\pm 0.02$ & $\bm{3.0\sigma}$ \\
            $\langle F_L\rangle^{[2,5]}$ & $0.81\pm 0.02$ & $0.68^{+0.16}_{-0.13}\pm 0.03$ & $0.8\sigma$ \\
            $\langle F_L\rangle^{[15,19]}$ & $0.34\pm 0.01$ & $0.29^{+0.07}_{-0.06}\pm 0.02$ & $0.6\sigma$ \\
            $\langle S_3\rangle^{[0.1,2]}$ & $0.02\pm 0.01$ & $-0.05^{+0.13}_{-0.13}\pm 0.01$ & $0.5\sigma$ \\
            $\langle S_3\rangle^{[2,5]}$ & $-0.01\pm 0.00$ & $-0.06^{+0.19}_{-0.23}\pm 0.01$ & $0.2\sigma$ \\
            $\langle S_3\rangle^{[15,19]}$ & $-0.21\pm 0.01$ & $-0.09^{+0.11}_{-0.12}\pm 0.01$ & $1.0\sigma$ \\
            $\langle S_4\rangle^{[0.1,2]}$ & $0.06\pm 0.01$ & $0.27^{+0.28}_{-0.18}\pm 0.01$ & $0.8\sigma$ \\
            $\langle S_4\rangle^{[2,5]}$ & $-0.15\pm 0.02$ & $-0.47^{+0.30}_{-0.44}\pm 0.01$ & $0.7\sigma$ \\
            $\langle S_4\rangle^{[15,19]}$ & $-0.30\pm 0.00$ & $-0.14^{+0.11}_{-0.11}\pm 0.01$ & $1.5\sigma$ \\
            $\langle S_7\rangle^{[0.1,2]}$ & $-0.02\pm 0.02$ & $0.04^{+0.12}_{-0.12}\pm 0.00$ & $0.5\sigma$ \\
            $\langle S_7\rangle^{[2,5]}$ & $-0.02\pm 0.04$ & $-0.03^{+0.18}_{-0.23}\pm 0.01$ & $0.0\sigma$ \\
            $\langle S_7\rangle^{[15,19]}$ & $0.00\pm 0.00$ & $0.13^{+0.11}_{-0.11}\pm 0.01$ & $1.2\sigma$ \\
            \hline
		\end{tabular}
		\caption{Angular observables of $B_s^0\to \phi\mu^+\mu^-$ included in the global fit.}
		\label{tab:data_B0_exp9} 
	\end{center}
\end{table}

\begin{table}[p]
	\begin{center}\footnotesize
		\begin{tabular}{l|c|c|c}
			\hline\hline
			\multicolumn{4}{c}{$\Lambda_b^0\to \Lambda\mu^+\mu^-$ differential branching ratio} \\
		    \hline\hline
            Observable & SM prediction & Experimental value & Pull\\
            \hline
			\multicolumn{4}{c}{LHCb\cite{Aaij:2015xza}}\\
		    \hline
            $10^7\times\langle \frac{d\textrm{BR}}{dq^2}\rangle^{[1,6]}$ & $0.10\pm 0.06$ & $0.09^{+0.06}_{-0.05}\pm 0.02$ & $0.2\sigma$ \\
            $10^7\times\langle \frac{d\textrm{BR}}{dq^2}\rangle^{[15,20]}$ & $0.71\pm 0.08$ & $1.20^{+0.09}_{-0.09}\pm 0.25$ & $1.7\sigma$ \\
            \hline
		\end{tabular}
		\caption{Binned differential branching ratio of $\Lambda_b^0\to \Lambda\mu^+\mu^-$ included in the global fit.}
		\label{tab:data_B0_exp10} 
	\end{center}
\end{table}

\begin{table}[p]
	\begin{center}\footnotesize
		\begin{tabular}{l|c|c|c}
			\hline\hline
			\multicolumn{4}{c}{$\Lambda_b^0\to \Lambda\mu^+\mu^-$ angular asymmetries} \\
		    \hline\hline
            Observable & SM prediction & Experimental value & Pull\\
            \hline
			\multicolumn{4}{c}{LHCb\cite{Aaij:2018gwm}}\\
		    \hline
            $\langle A^l_{FB} \rangle^{[15,20]}$ & $-0.36\pm 0.02$ & $-0.39\pm 0.04\pm 0.01$ & $0.8\sigma$ \\
            $\langle A^h_{FB}\rangle^{[15,20]}$ & $-0.27\pm 0.01$ & $-0.30\pm 0.05\pm 0.02$ & $0.5\sigma$ \\
            $\langle A^{lh}_{FB}\rangle^{[15,20]}$ & $0.14\pm 0.01$ & $0.25\pm 0.04\pm 0.01$ & $\bm{2.7\sigma}$ \\
            \hline
		\end{tabular}
		\caption{Angular observables of $\Lambda_b^0\to \Lambda\mu^+\mu^-$ included in the global fit.}
		\label{tab:data_B0_exp11} 
	\end{center}
\end{table}

\begin{table}[p]
	\begin{center}\footnotesize
		\begin{tabular}{l|c|c|c}
			\hline\hline
			\multicolumn{4}{c}{$B^+\to K^+\mu^+\mu^-$ angular observables} \\
		    \hline\hline
            Observable & SM prediction & Experimental value & Pull\\
            \hline
			\multicolumn{4}{c}{CMS\cite{Sirunyan:2018jll}}\\
		    \hline
            $\langle A_{FB} \rangle^{[1,6]}$ & $0.00\pm 0.00$ & $-0.14^{+0.07}_{-0.06}\pm 0.03$ & $1.8\sigma$ \\
            $\langle F_H\rangle^{[2,4.3]}$ & $0.02\pm 0.00$ & $0.85^{+0.34}_{-0.31}\pm 0.14$ & $\bm{2.2\sigma}$ \\
            \hline
		\end{tabular}
		\caption{Angular observables of $B^+\to K^+\mu^+\mu^-$ included in the global fit.}
		\label{tab:data_B0_exp12} 
	\end{center}
\end{table}

\begin{table}[p]
	\begin{center}\footnotesize
		\begin{tabular}{l|c|c|c}
			\hline\hline 
			\multicolumn{4}{c}{$B^0\to K^{*0}e^+e^-$ angular observables} \\
		    \hline\hline
            Observable & SM prediction & Experimental value & Pull\\
			\hline
			\multicolumn{4}{c}{Belle\cite{Wehle:2016yoi}}\\
            \hline
            $\langle P_4^{'}\rangle^{[1,4]}$ & $-0.01\pm 0.03$ & $0.34^{+0.41}_{-0.45}\pm 0.11$ & $0.8\sigma$ \\
            $\langle P_4^{'}\rangle^{[14.18,19]}$ & $-0.63\pm 0.01$ & $-0.15^{+0.41}_{-0.40}\pm 0.04$ & $1.2\sigma$ \\
            $\langle P_5^{'}\rangle^{[1,4]}$ & $0.17\pm 0.06$ & $0.51^{+0.39}_{-0.46}\pm 0.09$ & $0.8\sigma$ \\
            $\langle P_5^{'}\rangle^{[14.18,19]}$ & $-0.62\pm 0.03$ & $-0.91^{+0.36}_{-0.30}\pm 0.03$ & $0.9\sigma$ \\
			\hline
			\multicolumn{4}{c}{LHCb\cite{Aaij:2015dea}}\\
            \hline
            $\langle F_L\rangle^{[0.002,1.12]}$ & $0.18\pm 0.04$ & $0.16\pm 0.06\pm 0.03$ & $0.3\sigma$ \\
            \hline
		\end{tabular}
		\caption{Angular observables of $B^0\to K^{*0}e^+e^-$ included in the global fit.}
		\label{tab:data_B0_exp14} 
	\end{center}
\end{table}

\begin{table}[p]
	\begin{center}\footnotesize
		\begin{tabular}{l|c|c|c}
			\hline\hline 
			\multicolumn{4}{c}{$B^0\to K^{*0}e^+e^-$ differential branching ratio} \\
		    \hline\hline
            Observable & SM prediction & Experimental value & Pull\\
			\hline
			\multicolumn{4}{c}{LHCb\cite{Aaij:2013hha}}\\
            \hline
            $10^7\times\langle \frac{d\textrm{BR}}{dq^2}\rangle^{[0.003,1]}$ & $2.5\pm 0.4$ & $3.1^{+0.9}_{-0.9}\pm 0.2$ & $0.6\sigma$ \\
            \hline
		\end{tabular}
		\caption{Binned differential branching ratio of $B^0\to K^{*0}e^+e^-$ included in the global fit.}
		\label{tab:data_B0_exp15} 
	\end{center}
\end{table}

\begin{table}[p]
	\begin{center}\footnotesize
		\begin{tabular}{l|c|c|c}
			\hline\hline 
			\multicolumn{4}{c}{$B^+\to K^+e^+e^-$ differential branching ratio} \\
		    \hline\hline
            Observable & SM prediction & Experimental value & Pull\\
            \hline
			\multicolumn{4}{c}{LHCb\cite{Aaij:2014ora}}\\
            \hline
            $10^7\times\langle \frac{d\textrm{BR}}{dq^2}\rangle^{[1,6]}$ & $0.349\pm 0.059$ & $0.312^{+0.040}_{-0.031}$ & $0.5\sigma$ \\
            \hline
		\end{tabular}
		\caption{Binned differential branching ratio of $B^+\to K^{+}e^+e^-$ included in the global fit.}
		\label{tab:data_B0_exp17} 
	\end{center}
\end{table}

\begin{table}[h]
	\begin{center}\footnotesize
		\begin{tabular}{l|c|c|c}
			\hline\hline
			\multicolumn{4}{c}{$B\to X_sl^+l^-$ branching ratio} \\
		    \hline\hline
            Observable & SM prediction & Experimental value & Pull\\
            \hline
			\multicolumn{4}{c}{BaBar\cite{Lees:2013nxa}}\\
		    \hline
            $10^6\times\langle \textrm{BR}\rangle(B\to X_s\mu^+\mu^-)^{[1,6]}$ & $1.68\pm 0.17$ & $0.66^{+0.87}_{-0.80}\pm 0.07$ & $1.1\sigma$ \\
            $10^6\times\langle \textrm{BR}\rangle(B\to X_s\mu^+\mu^-)^{[14.2,25]}$ & $0.34\pm 0.04$ & $0.60^{+0.31}_{-0.29}\pm 0.00$ & $0.8\sigma$ \\
            $10^6\times\langle \textrm{BR}\rangle(B\to X_se^+e^-)^{[1,6]}$ & $1.74\pm 0.18$ & $1.93^{+0.51}_{-0.48}\pm 0.18$ & $0.3\sigma$ \\
            $10^6\times\langle \textrm{BR}\rangle(B\to X_se^+e^-)^{[14.2,25]}$ & $0.29\pm 0.04$ & $0.56^{+0.19}_{-0.18}\pm 0.00$ & $1.5\sigma$ \\
            \hline
		\end{tabular}
		\caption{Binned $B\to X_sl^+l^-$ branching ratio included in the global fit.}
		\label{tab:data_B0_exp13} 
	\end{center}
\end{table}

\begin{table}[h]
    \begin{center}\footnotesize
        \begin{tabular}{l|c|c|c}
            \hline\hline
            \multicolumn{4}{c}{$B_s^0\to \mu^+\mu^-$ branching ratio} \\
            \hline\hline
            Observable & SM prediction & Experimental value & Pull\\
            \hline
            \multicolumn{4}{c}{LHCb+CMS\cite{CMS:2014xfa}}\\
            \hline
            $10^9\times\overline{\textrm{BR}}(B_s^0\to \mu^+\mu^-)$ & $3.67\pm 0.15$ & $2.80^{+0.70}_{-0.60}$ & $1.2\sigma$ \\
            \hline
        \end{tabular}
        \caption{$B_s^0\to \mu^+\mu^-$ branching ratio included in the global fit.}
        \label{tab:data_B0_exp18} 
    \end{center}
\end{table}

\bibliographystyle{utphysmcite}
\bibliography{mybib}

\end{document}